\documentstyle{article}

\begin{document}
\setcounter{page}{1}
\title{{\bf INTRODUCTION A LA COSMOLOGIE}}
\author{J\'ulio C\'esar Fabris \\
 \mbox{\small Departamento de F\'{\i}sica} \\
 \mbox{\small Universidade Federal do
Esp\'{\i}rito Santo} \\
 \mbox{\small Goiabeiras - Vit\'oria - CEP29060-900} \\
 \mbox{\small Esp\'{\i}rito Santo - Br\'esil}}
\date{} \maketitle
 \begin{center}
 {\bf Ecole Internationale sur les Structures
G\'eom\'etriques et Applications\\
17-26 Novembre 2004\\ Dakar, S\'en\'egal}
\end{center}

\section{L'objet de la cosmologie}

La cosmologie a pour but l'\'etude de l'Univers comme un tout. De cette mani\`ere, elle
se distingue de l'astrophysique qui vise plut\^ot l'\'etude des objets particuliers qui
existent dans l'Univers, comme les \'etoiles, les galaxies, etc.  La cosmologie essaye de
r\'epondre aux questions du genre:
\begin{enumerate}
\item L'Univers, est-il statique ou \' evolue-t-il avec le temps? \item
Est-il globalement homog\`ene ou ses propriet\'e d\'ependent de la
position dans l'espace? \item Comment peut on expliquer la
formation des galaxies, amas de galaxies, et toutes les autres
structures que l'on observe? \item Comment se forment les
\'elements chimiques qui existent dans la nature? \item Quelle est
la nature des composants mat\'eriels qui remplissent l'Univers?
\end{enumerate}
Evidemment, cette liste n'\'epuise pas l'ensemble de probl\`emes
abord\'es par la cosmologie.
\par
La force dominant dans l'Univers \`a grande \'echelle est la
gravitation. Ainsi, on doit avoir une th\'eorie de la gravitation
pour essayer de d\'ecrire la structure et l'\'evolution de
l'Univers. Nous avons actuellement une th\'eorie de la gravitation
qui r\'epond de mani\`ere positive aux impositions th\'eoriques et
obsevationnelles: c'est la Relativit\'e G\'en\'erale. Du point de
vue th\'eorique elle incorpore les principes relativistes, ainsi
comme le principe d'\'equivalence; du point de vue
observationnele, la Relativit\'e G\'en\'erale a resist\'e aux
tests locaux auxquels elle a \'et\'e soumise. La Relativit\'e
G\'en\'erale d\'ecrit la gravitation comme la structure de
l'espace-temps cr\'e\'ee par la distribution de mati\`ere.
\par
Nous d\'ebuterons cet expos\'e sur la cosmologie en rappellant les
grandes lignes de la Relativit\'e G\'en\'erale.
L'Univers semble \^etre \`a grande \'echelle homog\`ene et
isotrope. Donc, ensuite, nous obtiendrons la structure
g\'eom\'etrique qui doit servir pour la description de cet Univers
homog\`ene et isotrope. Les possibles solutions dynamiques seront
determin\'ees. Les principaux param\`etres observationnels seront
d\'ecrits, ce qui permettra par la suite d'\'etablir le mod\`ele
cosmologique standard, celui qui fournit une description de
l'\'evolution de l'Univers en accord raisonable avec ce qu'on
observe.
\par
Les probl\`emes qui existent dans ce mod\`ele standard seront
pr\'esent\'es, ainsi que leur possibles solutions,
particuli\`erement le mod\`ele d'inflation. Nous finissons ces
expos\'es en d\'ecrivant certains probl\`emes ouverts aujourd'hui en
cosmologie et quelques perspectives futures.
\par
Ces expos\'es visent un public des math\'ematiciens et physiciens non
sp\'ecialis\'es dans le domaine de la cosmologie. Des r\'ef\'erences dans
ce domaine peuvent \^etre trouv\'ees en \cite{weinberg,bernstein,rich}.

\section{La Gravitation comme g\'eom\'etrie de l'espace-temps}

La premi\`ere th\'eorie de la gravitation a \'et\'e formul\'ee par
Newton pendant le XVII si\`ecle. Elle dit que la force
gravitationnelle entre deux corps ponctuels de masses $m_1$ et
$m_2$ est donn\'ee par l'expression
\begin{equation}
F = G\frac{m_1m_2}{r^2} \quad ,
\end{equation}
o\`u $G = 6,67\times10^{-11}\;N\,kg^{-2}\,m^{2}$ est la constante
gravitationnelle. La th\'eorie newtonienne con\c{c}oit la
gravitation comme une force qui agit instantan\'ement \`a distance.
En plus, si nous combinons la loi de force gravitationnelle
newtonienne avec la deuxi\`eme loi de Newton, nous obtenons pour
la force sur le corps de masse $m_2$, par exemple,
\begin{equation}
m_2a = G\frac{m_1m_2}{r^2} \quad \rightarrow \quad a =
G\frac{m_1}{r^2} \quad .
\end{equation}
Cela veut dire que tous les corps subissent la m\^eme
acc\'el\'eration due au corps $m_1$, ind\'ependamment de leur
masse.
\par
Cette derni\`ere propriet\'e est d\^ue au fait qu'on a
consid\`er\'e que la masse qui appara\^{\i}t dans la deuxi\`eme
loi de Newton est la m\^eme qui appara\^{\i}t dans la loi de force
gravitationnelle. Or, ceci du point de vue th\'eorique n'est pas
vrai: la masse qui appara\^{\i}t dans la deuxi\`eme loi est
reli\'ee aux propriet\'es inertieles du corps, c'est-\`a-dire,
\`a la tendance du corps \`a se maintenir dans un certain \'etat de
mouvement; la masse qui appara\^{\i}t dans la loi de
gravitation est une mesure de l'intensit\'e du champ
gravitationnel cr\'e\'e par le corps, \'etant par cons\'equent une
sorte de "charge gravitationnelle". La signification physique des
deux masses est donc compl\'etement differente, et leur
\'egalit\'e ressort du fait experimental que tous les corps
subissent la m\^eme acc\'eleration dans un champ gravitationnel.
Il n'y a aucune raison th\'eorique pour que \c{c}a soit le cas.
L'\'egalit\'e des masses inertielle et gravitationnelle est connue
comme le "principe d'equivalence".
\par
D'une mani\`ere plus pr\'ecise, dans le cas o\`u le contenu
mat\'eriel est vu comme un fluide, la gravitation newtonienne
conduit \`a un ensemble d'\'equations qui peuvent \^etre
appliqu\'ees \`a la description d'un mod\`ele cosmologique. La
premi\`ere concerne la conservation de la masse. Consid\'erons un
volume $V$ d\'efini par une surface ferm\'ee $S$. La variation de
la masse \`a l'interieur du volume est \'egale au courant de masse
\`a travers la surface:
\begin{eqnarray}
\frac{dM}{dt} = - \int_S \vec j.d\vec S \quad &\rightarrow& \quad
\frac{d}{dt}\int\rho dV = - \int\nabla.\vec j dV \quad \rightarrow
\quad \nonumber\\
\frac{\partial\rho}{\partial t} + \nabla\vec j = 0 \quad &,& \quad
\vec j = \rho\vec v.
\end{eqnarray}
\par
Consid\'erons maintenant la deuxi\`eme loi de Newton pour un
\'el\'ement de ce fluide de densit\'e $\rho$. Sur cet \'el\'ement
agissent le grandient de la pression et la force gravitationnelle:
\begin{eqnarray}
\rho\frac{d\vec v}{dt} = - \nabla p - \rho\nabla\phi \rightarrow
\nonumber\\
\frac{\partial\vec v}{\partial t} + \vec v.\nabla \vec v = -
\frac{\nabla p}{\rho} - \nabla\phi \quad ,
\end{eqnarray}
o\`u $\phi$ est le potentiel gravitationnel.
\par
Finalement, le potentiel gravitationnel d\^u \`a une distribution
de masse s'\'ecrit comme
\begin{equation}
\phi(\vec r) = - G\int \frac{\rho(\vec r')}{|\vec r - \vec r'|}dV
\quad .
\end{equation}
Utilisant le fait que
\begin{equation}
\nabla^2\frac{1}{|\vec r - \vec r'|} = - 4\pi\delta(\vec r - \vec
r') \quad ,
\end{equation}
nous obtenons
\begin{equation}
\nabla^2\phi = 4\pi G\rho \quad .
\end{equation}
\par
Ainsi, le syst\`eme d'\'equations convenable pour d\'ecrire une
cosmologie newtonienne est:
\begin{eqnarray}
\label{cn1}
\frac{\partial\rho}{\partial t} + \nabla.(\rho\vec v) = 0 \quad ,\\
\label{cn2} \frac{\vec v}{\partial t} + \vec v.\nabla \vec v =
- \frac{\nabla p}{\rho} - \nabla\phi \quad ,\\
\label{cn3} \nabla^2\phi = 4\pi G\rho \quad .
\end{eqnarray}
La premi\`ere \'equation est l'\'equation de continuit\'e, la
deuxi\`eme l'\'equation d'Euler et la troisi\`eme l'\'equation de
Poisson. Dans ces expressions, $\rho$ est la densit\'e d'un
fluide, $\vec v$ le champ de vitesse du fluide, $p$ la pression et
$\phi$ le potentiel gravitationnel.
\par
Malgr\'e son succ\`es \'evident, la th\'eorie de Newton contient
deux inconv\'enients majeurs du point de vue th\'eorique:
\begin{enumerate}
\item La propagation instantan\'ee de l'interaction
gravitationnelle est en flagrante contradiction avec les principes
de la Relativit\'e Restreinte qui \'etablissent une vitesse limite
dans la nature, celle de la lumi\`ere; \item Le principe d'\'equivalence n'est pas naturellement contenu dans la th\'eorie.
\end{enumerate}
\par
Ces probl\'emes ont \'et\'e abord\'es par Einstein, qui a
propos\'e une nouvelle th\'eorie de la gravitation. Dans cette
th\'eorie, la gravitation n'est plus vue comme une force, mais
comme la structure de l'espace-temps quadri-dimensionel. Ainsi,
les principes relativistes sont inclus, puisque nous avons
maintenant \`a faire \`a un espace-temps, et aussi le principe d'\'equivalence, puisque les corps se d\'eplacent tous de la m\^eme
mani\`ere dans une g\'eom\'etrie donn\'ee. L'id\'ee centrale de la
th\'eorie einsteinienne c'est que la g\'eom\'etrie de
l'espace-temps n'est pas une donn\'ee a priori, mais qu'elle est
d\'etermin\'ee par la distribution de mati\`ere. La relation
fondamentale est du genre,
\begin{equation}
\mbox{g\'eom\'etrie} = \mbox{mati\`ere} \quad .
\end{equation}
\par
Math\'ematiquement, nous avons que la g\'eom\'etrie sera donn\'ee
par les coefficients m\'etriques $g_{\mu\nu}$ de l'intervalle
spatio-temporel
\begin{equation}
ds^2 = g_{\mu\nu}dx^\mu\,dx^\nu \quad .
\end{equation}
La courbure d'une vari\'et\'e est caracteris\'e par le tenseur de
Riemann
\begin{equation}
R^\lambda_{\mu\gamma\nu} = \partial_\gamma\Gamma^\lambda_{\mu\nu}
-
\partial_\nu\Gamma^\lambda_{\mu\gamma} +
\Gamma^\rho_{\mu\nu}\Gamma^\lambda_{\gamma\rho} -
\Gamma^\rho_{\mu\gamma}\Gamma^\lambda_{\rho\nu} \quad .
\end{equation}
o\`u les symboles de Christofell s'\'ecrivent comme
\begin{equation}
\Gamma^\lambda_{\mu\nu} =
\frac{1}{2}g^{\lambda\gamma}\biggr(\partial_\mu g_{\gamma\nu} +
\partial_\nu g_{\gamma\mu} - \partial_\gamma g_{\mu\nu}\biggl)
\quad .
\end{equation}
Le tenseur de Riemann est connect\'e \`a la non-commutativit\'e
des d\'eriv\'ees covariantes:
\begin{equation}
V^\lambda_{;\mu;\nu} - V^\lambda_{;\nu;\mu} =
R^\lambda_{\gamma\mu\nu}V^\gamma \quad ,
\end{equation}
o\`u le point-virgule indique la d\'eriv\'ee covariante.
\par
A partir du tenseur de Riemann nous pouvons obtenir le tenseur de
Ricci $R_{\mu\nu}$ d\'efini comme
\begin{eqnarray}
& &R_{\mu\nu} = R^\lambda_{\mu\lambda\nu} = \nonumber\\
& &\partial_\lambda\Gamma^\lambda_{\mu\nu} -
\partial_\nu\Gamma^\lambda_{\mu\lambda} +
\Gamma^\lambda_{\mu\nu}\Gamma^\rho_{\lambda\rho} -
\Gamma^\lambda_{\mu\rho}\Gamma^\rho_{\lambda\nu} \quad ,
\end{eqnarray}
A partir du tenseur de Ricci, on construit le scalaire de Ricci
au moyen de l'expression
\begin{equation}
R = g^{\mu\nu}R_{\mu\nu} \quad .
\end{equation}
\par
Les coefficients m\'etriques seront solution des \'equations
dynamiques qui relient la g\'eom\'etrie \`a la distribution de
mati\`ere:
\begin{equation}
R_{\mu\nu} - \frac{1}{2}g_{\mu\nu}R = 8\pi GT_{\mu\nu} \quad ,
\end{equation}
o\`u $T_{\mu\nu}$ est le tenseur d'impulsion-energie. Nous avons
d\'ecrit auparavant le contenu du cot\'e gauche de l'\'equation
d'Einstein, c'est-\`a-dire, de sa partie geom\'etrique.
\par
Maintenant, il faut sp\'ecifier le contenu materiel. Ceci est
donn\'ee par le cot\'e droit de l'\'equation d'Einstein. En
g\'eneral, lorsque dans le fluide il n'existe pas de cisaillement
ni transfert de chaleur, le tenseur d'\'energie-impulsion s'\'ecrit
comme
\begin{equation}
T^{\mu\nu} = (\rho + p)u^\mu u^\nu - pg^{\mu\nu} \quad ,
\end{equation}
o\`u $u^\mu$ est la quadri-vitesse du fluide de densit\'e $\rho$
et pression $p$. Ces \'equations peuvent \^etre obtenues \`a
partir de la densit\'e lagrangienne
\begin{equation}
{\cal L} = \frac{1}{16\pi G}\sqrt{-g}R + {\cal L}_m \quad ,
\end{equation}
o\`u ${\cal L}_m$ est le lagrangien qui d\'ecrit la mati\`ere.
\par
 Lorsque nous avons
affaire \`a un espace-temps qui admet un feuilletage de telle
mani\`ere que la vari\'et\'e quadri-dimensionnelle $V^4$ prenne la
forme $V^4 = R\times S^3$, o\`u $S^3$ est une hypersurface
spatiale et $R$ d\'esigne la ligne temporelle, nous pouvons choisir
un syst\`eme de coordonn\'ees co-mobile, tel que la quadri-vitesse
prend la forme
\begin{equation}
u^\mu = (1,\vec 0) \quad .
\end{equation}
Avec ce syst\`eme de coordonn\'ees co-mobile, les composantes du
tenseur d'\'energie-impulsion s'\'ecrivent,
\begin{equation}
T^{00} = \rho \quad , \quad T^{ij} = - pg^{ij} \quad , \quad
T^{0i} = 0 \quad .
\end{equation}
Les \'equations d'Einstein se r\'eduisent \`a l'\'equation de Poisson
dans l'approximation du champ faible.
\par
Le tenseur de Riemann satisfait aux identit\'es,
\begin{equation}
R_{\mu\nu\lambda\gamma;\rho} + R_{\mu\nu\rho\lambda;\gamma} +
R_{\mu\nu\gamma\rho;\lambda} = 0 \quad ,
\end{equation}
qui sont connues comme {\it identit\'es de Bianchi}. Par
contraction, on obtient l'identit\'ee
\begin{equation}
\biggr(R^{\mu\nu} - \frac{1}{2}g^{\mu\nu}R\biggl)_{;\mu} = 0 \quad
,
\end{equation}
d'o\`u il suit, du fait des \'equations d'Einstein,
\begin{equation}
{T^{\mu\nu}}_{;\mu} = 0 \quad .
\end{equation}
Cette derni\`ere relation exprime la conservation du tenseur
d'impulsion-\'energie. Dans la limite newtonienne elle se r\'eduit
\`a la conservation de la masse, l'\'equation de la continuit\'e.
\par
Toutes ces relations doivent \^etre complet\'ees par l'\'equation
des geod\'esiques. Cette \'equation d\'ecrit comment une particule
test se d\'eplace dans un espace-temps donn\'e. L'\'equation des geod\'esiques est de la forme,
\begin{equation}
\frac{d^2x^\mu}{ds^2} +
\Gamma^\mu_{\lambda\gamma}\frac{dx^\lambda}{ds}\frac{dx^\gamma}{ds}
= 0 \quad .
\end{equation}
Dans la limite du champ faible, cette \'equation se reduit \`a
l'\'equation d'Euler, c'est-\`a-dire, la deuxi\`eme \'equation de
Newton \'ecrite pour un fluide.
\par
Avant de passer \`a l'\'etude des mod\`eles cosmologiques
relativistes, regardons ce que nous disent les \'equations
newtoniennes. A grande \'echelle, l'Univers est homog\`ene et
isotrope. Donc, la densit\'e, et par cons\'equent la pression, ne
peut d\'ependre que du temps, $\rho = \rho(t)$, $p = p(t)$. Ainsi,
nous en d\'eduisons, \`a partir de l'\'equation (\ref{cn1}), que
le champ de vitesse doit \^etre donn\'e par une fonction du temps
qui multiplie $\vec r$. Posons, par commodit\'e,
\begin{equation}
\vec v = \frac{\dot a}{a}\vec r \quad .
\end{equation}
Dans cette expression, $a$ doit \^etre une fonction uniquement du
temps. En plus, $\dot a = \frac{da}{dt}$. La raison pour laquelle on a mis l'expression pour la vitesse sous cette
forme deviendra plus claire par la suite. Ainsi, de (\ref{cn1}),
nous obtenons,
\begin{equation}
\dot\rho + 3\frac{\dot a}{a}\rho = 0 \quad \rightarrow \quad \rho
= \frac{\rho_0}{a^3} \quad , \quad \rho_0 = \mbox{constante} \quad
.
\end{equation}
Maintenant, consid\'erons l'\'equation de Poisson. Puisque le
cot\'e droit ne d\'epend que du temps, nous avons
\begin{equation}
\nabla\phi = \frac{4\pi G}{3}\rho\vec r \quad .
\end{equation}
Maintenant, en se rappellant que la pression ne depend que du
temps, nous trouvons pour l'\'equation d'Euler,
\begin{equation}
\frac{\ddot a}{a} = - \frac{4\pi G}{3}\rho = - \frac{4\pi
G}{3}\frac{\rho_0}{a^3} \quad .
\end{equation}
Cette \'equation admet l'int\'egrale premi\`ere,
\begin{equation}
\frac{\dot a^2}{2} - \frac{4\pi G}{3}\frac{\rho_0}{a} = - k \quad
.
\end{equation}
o\`u $k$ est une constante d'integration. Nous pouvons v\'erifier
que cette derni\`ere relation dit que l'\'energie cin\'etique par
unit\'e de masse, moins l'energie potentiel de masse, est \'egale
a l'\'energie par unit\'e de masse $k$. Lorsque $k > 0$, $a$
cro\^{\i}t jusqu'\`a atteindre une valeur maximale, apr\`es quoi
il d\'ecro\^{\i}t; si $k < 0$, $a$ peut cro\^{\i}tre ind\'efiniment;
le cas $k = 0$ correspond \`a la situation marginale. Nous
reviendrons \`a ce comportement lorsque nous traiterons du cas
relativiste. Observons que $a$ joue le r\^ole de "rayon de
l'Univers", quoique cette denomination est loin d'\^etre
pr\'ecise.
\par
Observons, finallement, que l'\'equation obtenue auparavant peut
\^etre re\'ecrite sous la forme,
\begin{equation}
3\biggr(\frac{\dot a}{a}\biggl)^2 + 3\frac{k}{a^2} = 8\pi G\rho
\quad .
\end{equation}
Nous retrouverons aussi cette \'equation, dite equation de
Friedmann, dans le cas relativiste. Mais, ceci ne veut pas dire
que le cas relativiste, en g\'eneral, est equivalent au cas
newtonien, comme on verra par la suite.

\section{La m\'etrique de l'espace-temps}

Dans la th\'eorie de la Relativit\'e G\'en\'erale, la gravitation
est exprim\'ee comme la g\'eom\'etrie de l'espace-temps \`a quatre
dimensions. C'est \`a dire, il faut d\'eterminer les coefficients
de la m\'etrique
\begin{equation}
ds^2 = g_{\mu\nu}dx^\mu dx^\nu \quad ,
\end{equation}
o\`u $\mu,\nu = 0,1,2,3$. Nous cherchons une m\'etrique qui
d\'ecrive un Univers qui soit globalement homog\`ene et isotrope
et qui \'evolue avec le temps. La construction de la m\'etrique qui
d\'ecrit un Univers ob\'eissant aux hypoth\`eses (soutenues par
les observations) d'homog\'en\'eit\'e et d'isotropie suit un
sch\'ema assez simple. Si l'Univers est homog\`ene et isotrope, la
section spatiale pour chaque temps $t$ fixe doit \^etre espace
homog\`ene. Donc, on peut \'ecrire la m\'etrique comme
\begin{equation}
ds^2 = dt^2 - g_{ij}dx^idx^j \quad ,
\end{equation}
o\`u $i,j = 1,2,3$. Le fait que le coefficient du terme temporel
soit \'egal \`a l'unit\'e d\'ecoule de l'emploi d'un syst\`eme de
coordonn\'ees co-mobile. La m\'etrique $g_{ij}$ d\'efinie sur la
section spatiale doit \^etre celle d'une sph\`ere, d'un espace
euclidien ou d'une pseudo-sph\`ere, dont l'\'echelle peut changer avec
le temps. Ce sont les trois cas d'espaces homog\`enes. La
m\'etrique ne peut prendre que la forme
\begin{equation}
ds^2 = dt^2 - g_{ij}dx^idx^j = dt^2 - a^2(t)\biggr[\frac{dr^2}{1
- kr^2} + r^2(d\theta^2 + \sin^2\theta d\phi^2)\biggl] \quad ,
\end{equation}
o\`u le param\`etre $k$ peut prendre les valeurs $1$ (sph\`ere),
$0$ (section euclidienne) et $k = - 1$ (pseudo-sph\`ere). Ces
valeurs sont obtenues apr\`es un r\'e-\'echellonage de la
courbure. La fonction $a(t)$ est nomm\'ee {\it facteur
d'\'echelle} puisqu'elle donne l'\'echelle de ces espaces
homog\`enes comme fonction du temps.
\par
Lorsque $k = 0$, cette forme de la m\'etrique est \'evidente,
puisque nous avons une section spatiale euclidienne. Pour les
autres cas, il faut calculer explicitement cette forme.
Consid\'erons, comme exemple, le cas de la sph\`ere. Pour
simplifier, prenons la sph\`ere bi-dimensionelle. La m\'etrique
prend la forme,
\begin{equation}
ds^2 = R^2(d\theta^2 + \sin^2\theta d\phi^2) \quad,
\end{equation}
o\`u $R$ est le rayon de la sph\`ere. Nous pouvons faire une
projection st\'er\'eographique de la sph\`ere sur le plan ($x,y$),
correspondant aux transformations
\begin{eqnarray}
x = 2R\mbox{cotan}\frac{\theta}{2}\cos\phi \quad ,\\
y = 2R\mbox{cotan}\frac{\theta}{2}\sin\phi \quad .
\end{eqnarray}
En utilisant ces transformations, nous obtenons les relations
suivantes:
\begin{eqnarray}
\sin\theta = \frac{2f}{1 + f^2} \quad ; \\
d\theta = \frac{1}{2R^2}\frac{1}{f}\frac{1}{1 + f^2}(xdx + ydy)
\quad ;\\
d\phi = \frac{1}{4R^2}\frac{1}{f^2}(xdy - ydx) \quad ;\\
f = \frac{1}{2R}\sqrt{x^2 + y^2} \quad .
\end{eqnarray}
Avec ces relations on obtient pour la m\'etrique de la sph\`ere
projet\'ee sur le plan euclidien,
\begin{equation}
\label{sp}
ds^2 = \frac{1}{\biggr[1 + \frac{x^2 +
y^2}{4R^2}\biggl]}(dx^2 + dy^2) \quad .
\end{equation}
Cette projection montre que la sph\`ere n'est pas localement
isom\'etrique au plan euclidien, \`a cause du facteur conforme en
(\ref{sp}). Le cas de la tri-sph\`ere conduit exactement au m\^eme
r\'esultat, pouvant \^etre \'ecrit en termes de coordonn\'ees
sph\'eriques ($r',\theta,\phi$) sur l'espace euclidien comme
\begin{equation}
\label{ms1}
ds^2 = \frac{1}{\biggr(1 +
\frac{kr'^2}{4}\biggl)^2}[dr'^2 + r'^2(d\theta^2 + \sin^2\theta
d\phi^2)] \quad ,
\end{equation}
o\`u nous avons d\'efini la courbure comme
\begin{equation}
k = \frac{1}{R^2} \quad .
\end{equation}
\par
Maintenant, nous introduisons la transformation
\begin{equation}
\frac{r'}{1 + \frac{r'^2}{4}} = r \quad \rightarrow\quad r' =
2\biggr(\frac{1 + \sqrt{1 - r^2}}{r}\biggl) \quad .
\end{equation}
Apr\`es quelques calculs, on trouve
\begin{equation}
\frac{dr'}{1 + \frac{r'^2}{4}} = - \frac{dr}{\sqrt{1 - r^2}} \quad
,
\end{equation}
et ins\'erant ces transformations dans (\ref{ms1}), on retrouve la
section spatiale de la m\'etrique de FRW pour le cas $k = 1$. Le
cas de la pseudo-sph\`ere suit les m\^emes lignes.
\par
Connaissant la m\'etrique, on peut \'evaluer le c\^ot\'e gauche
des \'equations d'Einstein. Pour ce faire, on doit d'abord
calculer les coefficients de Christoffel et ensuite le tenseur de
Ricci $R_{\mu\nu}$ et le scalaire de Ricci $R$. Nous \'ecrivons
donc la m\'etrique sous la forme
\begin{equation}
g_{00} = 1 \quad , \quad g_{ij} = a^2(t)\gamma_{ij} \quad ,
\end{equation}
o\`u $\gamma_{ij}$ est la m\'etrique \`a courbure constante sur la
section spatiale. Ainsi, les coefficients de Christoffel prennent
la forme
\begin{equation}
\Gamma^i_{oj} = \frac{\dot a}{a}\delta^i_ j \quad , \quad \Gamma^0_{ij} = a\dot a\gamma_{ij}
\quad , \quad
\Gamma^i_{jk} = \tilde\Gamma^i_{jk} \quad ,
\end{equation}
o\`u $\tilde\Gamma^i_{jk}$ est le coefficient de Christoffel
construit \`a partir de la m\'etrique de la section spatiale
$\gamma_{ij}$. Toutes les autres composantes du coefficient de
Christoffel sont nulles. Ainsi, on peut \'evaluer les composantes
du tenseur de Ricci. Nous trouvons:
\begin{eqnarray}
R_{00} &=& - 3\frac{\ddot a}{a} \quad ,\\
R_{ij} &=& - a\ddot a - 2\dot a^2 - 2k\quad .
\end{eqnarray}
Pour le scalaire de Ricci, il en r\'esulte
\begin{equation}
R = - 6\biggr[\frac{\ddot a}{a} + \biggr(\frac{\dot a}{a}\biggl)^2 + \frac{k}{a^2}\biggl]
\quad.
\end{equation}
Ainsi, les composantes non nulles du tenseur d'Einstein sont
\begin{eqnarray}
G_{00} &=& 3\biggr(\frac{\dot a}{a}\biggl)^2 + 3\frac{k}{a^2} \quad , \\
G_{ij} &=& - 2a\ddot a - \dot a^2 - k \quad .
\end{eqnarray}
\par
Maintenant il faut tourner le regard du c\^ot\'e droit des
\'equations d'Einstein. Le tenseur d'impulsion-\'energie s'\'ecrit
comme
\begin{equation}
T^{\mu\nu} = (\rho + p)u^\mu u^\nu - pg^{\mu\nu} \quad .
\end{equation}
Nous pouvons toujours trouver un syst\`eme de ref\'erentiel qui
suit le fluide. Dans ce r\'ef\'erentiel, les particules du fluide
sont au repos. Donc, la quadrivitesse s'\'ecrit comme
\begin{equation}
u^\mu = (1,0,0,0) \quad .
\end{equation}
Ainsi les composantes du tenseur d'impulsion-\'energie
s'\'ecrivent comme
\begin{eqnarray}
T^{00} &=& \rho  \quad ,\\
T^{ij} &=& p\,a^2\gamma^{ij} \quad .
\end{eqnarray}
De cette mani\`ere, les \'equations d'Einstein prennent la forme
\begin{eqnarray}
\label{ec1}
3\biggr(\frac{\dot a}{a}\biggl)^2 + 3\frac{k}{a^2} &=& 8\pi G\rho \quad , \\
\label{ec2}
2\frac{\ddot a}{a} + \biggr(\frac{\dot a}{a}\biggl)^2 + \frac{k}{a^2} &=& - 8\pi Gp \quad .
\end{eqnarray}
En plus de ces \'equations, nous avons la loi de conservation,
exprim\'ee par fait que la divergence du tenseur
d'\'energie-impulsion est nulle:
\begin{equation}
{T^{\mu\nu}}_{;\mu} = 0 \quad .
\end{equation}
En utilisant la m\'etrique de FRW, nous trouvons
\begin{equation}
\label{ec3}
\dot\rho + 3\frac{\dot a}{a}(\rho + p) = 0 \quad .
\end{equation}
\par
Remarquons, qu'au contraire du cas newtonien, la pression maintenant intervient
dans la loi de conservation relativiste. Cela conduit \`a des sc\'enarios distincts des cas newtoniens correspondants,
sauf quand la pression est nulle.
Les \'equations (\ref{ec1},\ref{ec2},\ref{ec3}) forment l'ensemble
des \'equations qui d\'ecrivent l'\'evolution d'un Univers
homog\`ene et isotrope. Le but est de trouver la fonction $a(t)$,
ainsi que $\rho(t)$.
\par
Tout d'abord, il faut remarquer que les trois \'equations
ci-dessus ne sont pas ind\'ependantes: elles sont reli\'ees par
les identit\'es de Bianchi. Donc, on ne peut utiliser que deux de
ces \'equations, au choix. Mais, dans ce cas, nous avons trois
fonctions inconnues ($a(t)$, $\rho(t)$ et $p(t)$) et deux
\'equations uniquement. Cela est assez naturel: les solutions de
ces \'equations ne sont pas les m\^emes suivant le type de
mati\`ere qui remplit l'Univers. Le type de fluide remplissant
l'Univers est caract\'eris\'e par {\it l'\'equation d'\'etat}, qui
\'etablit comment la pression d\'epend de la densit\'e, $p =
p(\rho)$.
\par
Dans les cas relativistes, la pression en g\'en\'eral est proportionnelle \`a la
densit\'e. Ainsi, nous pouvons supposer que
\begin{equation}
\label{ede}
p = \alpha\rho \quad ,
\end{equation}
o\`u $\alpha$ est un param\`etre, en principe constant. Trois cas
sont consid\'er\'es comme les plus importants en cosmologie:
\begin{itemize}
\item $\alpha = 0$, ce qui repr\'esente la mati\`ere sans
interaction, dite "poussi\`ere"; \item $\alpha = 1/3$, ce qui
repr\'esente un gaz de photons, dite "radiation"; \item $\alpha =
1$, ce qui repr\'esente la mati\`ere dans un \'etat hautement
condens\'e, dite "rigide";
\item $\alpha = - 1$, ce qui repr\'esente l'\'etat du vide quantique.
\end{itemize}
En fait, pour des raison qui seront discut\'ees plus tard, de plus en plus d'autres
\'equations d'\'etat sont prises en compte, supposant en particulier que
$\alpha < 0$.
\par
En possession de l'\'equation d'\'etat (\ref{ede}) nous pouvons ais\'ement int\'egrer
l'\'equation de conservation (\ref{ec3}). En fait, nous obtenons
\begin{equation}
\dot\rho + 3(1 + \alpha)\frac{\dot a}{a}\rho = 0 \quad \Rightarrow \quad \rho = \rho_0\,a^{- 3(1 + \alpha)}\quad,
\end{equation}
$\rho_0$ \'etant une constante d'int\'egration.
Ainsi, l'\'equation (\ref{ec1}) devient
\begin{equation}
\biggr(\frac{\dot a}{a}\biggl)^2 + \frac{k}{a^2} = \frac{8\pi G}{3}\frac{\rho_0}{a^{3(1 + \alpha)}} \quad .
\end{equation}
Cette \'equation admet une solution exacte sous la forme de
fonctions hyperg\'eom\'etrique. N\'eanmoins, il s'av\`ere plus
int\'eressant du point de vue physique de la r\'esoudre pour
certains cas particuliers qui, \`a la limite, sont les plus proche
de ce que l'on observe.
\par
Tout d'abord, il existe une solution g\'en\'erale lorsque la
courbure de l'Univers est nulle, $k = 0$. Dans ce cas, la section
spatiale est euclidienne (mais, l'espace quadri-dimensionel lui
m\^eme a une courbure). Dans ce cas,
\begin{equation}
\biggr(\frac{\dot a}{a}\biggl)^2 = \frac{8\pi G}{3}\frac{\rho_0}{a^{3(1 + \alpha)}}
\quad
\Rightarrow \quad a(t) = a_0 t^\frac{2}{3(1 + \alpha)} \quad .
\end{equation}
Plusieurs caract\'eristiques du mod\`ele cosmologique standard
peuvent \^etre d\'eduites de cette solution, qui est
remarquablement simple. Tout d'abord, elle montre un Univers en
expansion. En fait, le facteur d'\'echelle cro\^{\i}t avec le
temps. En remontant vers le pass\'e, on trouve un moment ($t = 0$
en l'occurrence), o\`u le facteur d'\'echelle s'annule. Cela
correspond \`a une singularit\'e, la singularit\'e initiale. Cela
correspond-il \`a une vraie singularit\'e? Ceci est une question
importante en relativit\'e g\'en\'erale puisque c'est une th\'eorie
invariante par des transformations arbitraires de coordonn\'ees:
il se peut qu'une singularit\'e apparaisse, mais que ce soit une
simple cons\'equence du choix de coordonn\'ees; dans un autre
syst\`eme de coordonn\'ees cette singularit\'e peut
dispara\^{\i}tre. Un exemple important o\`u cela arrive est le cas
des solutions \`a sym\'etries sph\'eriques: lorsqu'on utilise le
syst\`eme de coordonn\'ees sph\'eriques usuel, il appara\^{\i}t
deux singularit\'es. Mais, en fait, une des ces "singularit\'es"
est un simple effet du choix de coordonn\'ees, et dans d'autres
syst\`emes de coordonn\'ees elle dispara\^{\i}t, donnant lieu \`a
un comportement r\'egulier.
\par
Pour v\'erifier la vraie nature de la singularit\'e initiale, le
plus commode est d'\'evaluer les invariants de courbure: ce sont
des scalaires construits \`a partir du tenseur de Riemann, de
Ricci et du scalaire de Ricci. Puisque ce sont des scalaires, leur
valeur est ind\'ependante du choix de coordonn\'ees. \'Evaluons
l'invariant le plus simple, le scalaire de Ricci. Pour la
m\'etrique de FRW, il prend la forme
\begin{equation}
R = - 6 \biggr[\frac{\ddot a}{a} + \biggr(\frac{\dot a}{a}\biggl)^2 + \frac{k}{a^2}\biggl]
= - \frac{4}{3}\frac{1 - 3\alpha}{(1 + \alpha)^2}\frac{1}{t^2} \quad ,
\end{equation}
o\`u nous avons utilis\'e les solutions trouv\'ees auparavant pour
$k = 0$. On remarque donc que $R \rightarrow \infty$ lorsque $t
\rightarrow 0$. Cela veut dire que la courbure diverge \`a
l'origine: nous avons \`a faire \`a une vraie singularit\'e. On
reviendra sur cette question plus tard.
\par
Nous pouvons r\'esoudre les \'equations
(\ref{ec1},\ref{ec2},\ref{ec3}) lorsque la courbure spatiale est
diff\'erente de z\'ero. Tout d'abord, nous reprenons les \'equations
(\ref{ec1},\ref{ec3}), et nous les combinons pour avoir
\begin{equation}
\biggr(\frac{\dot a}{a}\biggl)^2 + \frac{k}{a^2} = \frac{8\pi G}{3}\frac{\rho_0}{a^{3(1+\alpha)}}
\quad .
\end{equation}
Maintenant, nous pouvons reparam\'etriser le facteur temporel en
d\'efinissant le temps conforme $\eta$ par l'expression
\begin{equation}
dt = ad\eta \quad \rightarrow \quad \eta = \int\frac{dt}{a(t)} \quad .
\end{equation}
Ainsi, l'\'equation de Friedmann devient
\begin{equation}
\biggr(\frac{a'}{a}\biggl)^2 + k = \frac{8\pi G}{3}\frac{\rho_0}{a^{1 + 3\alpha}} \quad .
\end{equation}
Nous d\'efinissons
\begin{equation}
M = \frac{8\pi G}{3}\rho_0 \quad .
\end{equation}
Ainsi, nous pouvons \'ecrire
\begin{equation}
\frac{a^\frac{- 1 + 3\alpha}{2}da}{\sqrt{M - k\,a^{1 + 3\alpha}}} = d\eta \quad .
\end{equation}
Choissant la nouvelle variable
\begin{equation}
u = a^\frac{1 + 3\alpha}{2} \quad \rightarrow \quad du = \frac{1 + 3\alpha}{2}a^\frac{- 1 + 3\alpha}{2} \quad ,
\end{equation}
l'int\'egrale peut \^etre \'ecrite sous la forme
\begin{equation}
\int \frac{du}{\sqrt{M - k\,u^2}} = \frac{1 + 3\alpha}{2}\eta \quad .
\end{equation}
Avec un nouveau changement de variable
\begin{equation}
u = \sqrt{M}v \quad ,
\end{equation}
nous aboutissons \`a
\begin{equation}
\label{mc1}
\int  \frac{dv}{\sqrt{1 - k\,v^2}} = \frac{1 + 3\alpha}{2}\eta \quad .
\end{equation}
\par
La solution de l'\'equation (\ref{mc1}) d\'epend de la valeur de
$k$. Pour $k = 1$, nous pouvons faire un nouveau changement de
variable
\begin{equation}
v = \sin\theta \quad ,
\end{equation}
conduisant au resultat
\begin{equation}
\label{s1}
v = \sin\biggr[\frac{1 + 3\alpha}{2}\eta\biggl] \quad \rightarrow \quad a = a_0\sin^\frac{2}{1 + 3\alpha}\biggr[\frac{1 + 3\alpha}{2}\eta\biggl] \quad .
\end{equation}
Pour $k = - 1$, nous choisissons,
\begin{equation}
v = \sinh\theta \quad ,
\end{equation}
et le facteur d'\'echelle prend la forme,
\begin{equation}
\label{s2}
a = a_0\sinh^\frac{2}{1 + 3\alpha}\biggr[\frac{1 + 3\alpha}{2}\eta\biggl] \quad .
\end{equation}
La solution (\ref{s1}) repr\'esente, pour $\alpha > -
\frac{1}{3}$, un Univers initialement en expansion, et qui entre
par la suite dans une phase de contraction. On l'appelle {\it
Univers ferm\'e}. D'autre part, la solution (\ref{s2})
repr\'esente, pour $\alpha > - \frac{1}{3}$ aussi, un Univers qui
est toujours en expansion. On l'appelle {\it Univers ouvert}.
\par
Pour $\alpha < - 1/3$, les solutions pour $k = - 1$ montrent un
Univers en contraction, qui atteint une singularit\'e. En fait,
cela correspond au m\^eme sc\'enario qu'un Univers en expansion si
on fait le changement $t \rightarrow - t$. D'autre part, lorsque
$k = 1$, nous avons un rebond: le facteur
d'\'echelle a initialement une valeur infinie, atteint au cours
de son \'evolution une valeur minimale pour atteindre de nouveau
une valeur infinie. Cela correspond \`a un Univers non singulier.
Pour chaque cycle, le temps cosmique est d\'efini dans
l'intervalle $- \infty < t < \infty$. Il faut remarquer que les
invariants de courbure restent toujours born\'es, ce qui confirme
l'impression qu'il n'existe pas de singularit\'e dans ce cas.

\section{Les param\`etres cosmologiques}

Un mod\`ele cosmologique peut \^etre caract\'eris\'e par un certain
nombre de param\`etres. Pour d\'efinir ces param\`etres, nous
revenons \`a l'\'equation de Friedmann,
\begin{equation}
3\biggr(\frac{\dot a}{a}\biggl)^2 + 3\frac{k}{a^2} = 8\pi G\sum_{i=1}^n \rho_i \quad ,
\end{equation}
o\`u les $\rho_i$ repr\'esentent les differents composantes mat\'erielles de l'Univers.
\c{C}a peut \^etre une combinaison de radiation ($p_r = \rho_r/3$), poussi\`ere
($p_m = 0$), etc. Chaque composante mat\'erielle ob\'eit \`a une \'equation de conservation:
\begin{equation}
\dot\rho_ i + 3\frac{\dot a}{a}(\rho_i + p_i) = 0 \quad .
\end{equation}
Si nous admettons que chacune de ces composantes ob\'eit \`a une \'equation d'\'etat barotropique,
$p_i = \alpha_i\rho_i$, nous obtenons \`a partir de l'\'equation de conservation,
\begin{equation}
\rho_ i = \frac{\rho_{i0}}{a^{3(1 + \alpha_i)}} \quad .
\end{equation}
Ainsi, l'\'equation de Friedmann prend la forme
\begin{equation}
\biggr(\frac{\dot a}{a}\biggl)^2 = \frac{8\pi}{3} G\sum_{i=1}^n
\frac{\rho_{i0}}{a^{3(1 + \alpha_i)}} - \frac{k}{a^2} \quad .
\end{equation}
\par
Nous d\'efinissons tout d'abord le param\`etre de Hubble:
\begin{equation}
H \equiv \frac{\dot a}{a} \quad .
\end{equation}
L'inverse de cette quantit\'e \'evalu\'ee \`a un moment donn\'e $t$ est proportionel \`a l'\^age
de l'Univers \`a ce moment:
\begin{equation}
T \propto \frac{1}{H} \quad .
\end{equation}
Par exemple, pour l'Univers spatialement plat ($k = 0$) rempli de mati\`ere incoh\'erente
($p = 0$), nous avons $a \propto t^\frac{2}{3}$, ce qui conduit \`a
\begin{equation}
T = \frac{3}{2}\frac{1}{H} \quad .
\end{equation}
L'int\'er\^et de cette quantit\'e vient du fait que c'est un observable, c'est-\`a-dire,
qu'elle est directement mesurable, comme on le verra par la suite.
\par
Une autre quantit\'e importante en cosmologie est le param\`etre de d\'ec\'el\'eration
$q$, d\'efini comme
\begin{equation}
\label{pd}
q = - \frac{a\ddot a}{\dot a^2} \quad .
\end{equation}
Remarquons que ce param\`etre d\'epend de la deuxi\`eme d\'eriv\'ee du facteur d'\'echelle.
Donc, il nous informe sur le changement de la vitesse d'expansion. Le signe moins dans le
deuxi\`eme membre de l'\'equation (\ref{pd}) indique que la valeur de ce param\`etre doit
\^etre positive si l'Univers est en train de d\'ec\'elerer, tandis qu'elle doit \^etre
n\'egative si l'Univers est en train d'acc\'el\'erer. Cela donne la  conviction
que l'expansion de l'Univers doit \^etre d\'ec\'el\'er\'ee, au vu de son contenu mat\'eriel.
Mais cette conviction s'est av\'er\'ee incorrecte plus r\'ecemment, comme nous verrons
plus tard.
\par
Les valeurs de la constante de Hubble et du param\`etre de
d\'ec\'el\'eration aujourd'hui sont denom\'ees $H_0$ et $q_0$. Nous
pouvons, d'autre part, choisir une \'echelle en imposant que le
facteur d'\'echelle soit \'egal \`a l'unit\'e aujourd'hui: $a(t_0)
= a_0 = 1$. La distance propre d'un certain point \`a une origine
choisie du syst\`eme de coordonn\'ees est donn\'ee par $r(t) =
a(t)r$. Avec ce choix d'\'echelle, la distance propre est \'egale
\`a la distance coordonn\'ee aujourd'hui. Ainsi, l'\'equation de
Friedmann s'\'ecrit aujourd'hui comme
\begin{equation}
H_0^2 = \frac{8\pi}{3} G\sum_{i=1}^n \rho_{i0} - k \quad ,
\end{equation}
o\`u les $\rho_{i0}$ sont les densit\'es des diff\'erentes
composantes aujourd'hui. Donc,
\begin{equation}
1 = \frac{8\pi}{3H_0^2}G \sum_{i=1}^n \rho_{i0} - \frac{k}{H_0^2}
\quad .
\end{equation}
Avec les d\'efinitions
\begin{equation}
\Omega_{i0} = \frac{8\pi}{3H_0^2}G\rho_{i0} \quad , \quad
\Omega_{k0} = - \frac{k}{H_0^2} \quad ,
\end{equation}
nous aboutissons \`a l'expression
\begin{equation}
1 = \sum_{i=1}^n\Omega_{i0} + \Omega_{k0} \quad .
\end{equation}
Lorsque $\Omega_{k0} = 0$, nous obtenons la densit\'e critique
\begin{equation}
\Omega_c = \sum_{i=1}^n\Omega_{i0} = 1 \quad .
\end{equation}
La densit\'e critique est la valeur de la densit\'e totale pour
laquelle la section spatiale est plate.
\par
Utilisant la composante $ii$ des \'equations d'Einstein, nous
avons
\begin{equation}
2\frac{\ddot a}{a} + \biggr(\frac{\dot a}{a}\biggl)^2 +
\frac{k}{a^2} = - 8\pi G\sum_{i=1}^n\,p_i \quad ,
\end{equation}
o\`u $p_i$ est la pression associ\'e \`a la composante $i$ du
contenu mat\'eriel de l'Univers. Combinant cette \'equation avec
l'\'equation de Friedmann, nous obtenons,
\begin{equation}
\frac{\ddot a}{a} = - \frac{4\pi}{3}G\sum_{i=1}^n\biggr(\rho_i +
3p_i\biggl) \quad .
\end{equation}
Supposons maintenant que la pression de chaque composante d\'epende
lin\'eairement de la densit\'e:
\begin{equation}
p_i = \alpha_i\rho_i \quad ,
\end{equation}
o\`u les $\alpha_i$ sont des constantes. En exprimant
l'\'equation r\'esultante en termes des quantit\'es evalu\'ees
aujourd'hui, nous obtenons
\begin{equation}
q_0 = - \frac{1}{2}\sum _{i=1}^n(1 + 3\alpha_i)\Omega_{i0} \quad .
\end{equation}
Donc, comme toutes les formes de mati\`ere ont, en principe, une pression
positive, le param\`etre de d\'ec\'eleration devrait \^etre positif: la vitesse d'expansion de l'Univers
d\'ecro\^{\i}t.
\par
Une donn\'ee observationnelle tr\`es importante est le
d\'ecalage vers le rouge $z$. Elle est directement en rapport avec la
vitesse avec laquelle un objet s'\'eloigne de nous et, par la loi
de Hubble, avec la distance entre cet objet et nous. Pour
d\'efinir cette quantit\'e, consid\'erons une source (par exemple,
une galaxie) qui est \`a une distance $d$ de nous, dor\'enavant
{\it les observateurs}. Prenons le d\'eplacement radial d'un
photon.
\begin{equation}
ds^2 = c^2dt^2 - a^2(t)dr^2 = 0 \quad .
\end{equation}
Le photon est \'emis \`a un temps $t_e$ (temps d'\'emission) par
une source \`a une distance $r$ de l'observateur, et re\c{cu} par
ce m\^eme observateur \`a un temps $t_r$ (temps de r\'eception).
Nous avons donc,
\begin{equation}
\int_r^0 dr' = c\int_{t_e} ^{t_r}\frac{dt}{a(t)}  \rightarrow r =
c\int_{t_r} ^{t_e}\frac{dt}{a(t)}
\end{equation}
Mais, nous devons nous rappeller que la lumi\`ere, classiquement, est {\it un
train d'onde} qui n'est pas \'emis instantan\'ement. Ainsi, le
photon commence \`a \^etre \'emis au temps $t_e$ et finit
d'\^etre \'emis au temps $t_e + \delta_e$; il commence \`a
\^etre re\c{c}u par l'observateur dans un temps $t_r$ et finit
d'\^etre re\c{c}u dans un temps $t_r + \delta_r$. Comme la source
et l'observateur ont des coordonn\'ees co-mobiles fixes, nous
avons
\begin{equation}
\int_r^0 dr' = c\int_{t_e + \delta_e} ^{t_r +
\delta_r}\frac{dt}{a(t)} \rightarrow r = c\int_{t_r + \delta_r}
^{t_e + \delta_e}\frac{dt}{a(t)}
\end{equation}
Ainsi,
\begin{equation}
\int_{t_e} ^{t_r}\frac{dt}{a(t)} = \int_{t_e + \delta_e} ^{t_r +
\delta}\frac{dt}{a(t)} \rightarrow \int_{t_e} ^{t_e +
\delta}\frac{dt}{a(t)} = \int_{t_r} ^{t_r +
\delta_r}\frac{dt}{a(t)}
\end{equation}
En g\'eneral, $\delta$ est un intervale de temps tr\`es petit.
Pour les fr\'equences de la lumi\`ere visible, la p\'eriode de la
radiation est de l'ordre de $10^{-15}\,s$. Ainsi, nous pouvons
utiliser l'approximation
\begin{equation}
\frac{\delta_e}{a(t_e)} \approx \frac{\delta_r}{a(t_r)} \quad
\rightarrow \quad \frac{\delta_r}{\delta_e} \approx
\frac{a(t_r)}{a(t_e)} \quad .
\end{equation}
La relation entre la p\'eriode et la longueur d'onde est
\begin{equation}
T = \frac{\lambda}{c} \quad .
\end{equation}
Donc, nous pouvons \'ecrire
\begin{equation}
\frac{\lambda_r}{\lambda_e} \approx \frac{a_r}{a_e} \quad .
\end{equation}
D\'efinissons
\begin{equation}
z = \frac{\lambda_r - \lambda_e}{\lambda_e} =
\frac{\lambda_r}{\lambda_e} - 1 \quad .
\end{equation}
Ainsi,
\begin{equation}
z + 1 = \frac{a_r}{a_e} \quad .
\end{equation}
En choisissant le temps de r\'eception aujourd'hui et en fixant
l'\'echelle telle que $a_0 = a_r = 1$, nous obtenons finalement
\begin{equation}
z = - 1 + \frac{1}{a(t)} \quad.
\end{equation}
\par
Consid\'erons maintenant le facteur d'\'echelle $a(t)$. Faisons un
d\'evelopement limit\'e autour du temps aujourd'hui $t_0$. Nous
obtenons
\begin{eqnarray}
a(t) = a(t_0) + \dot a(t)\biggr|_{t = t_0}(t - t_0) +
\frac{1}{2}\ddot a(t)\biggr|_{t = t_0}(t - t_0)^2 + ... \nonumber
\\
= a(t_0)\biggr[1 + \frac{\dot a_0}{a_0}(t - t_0) +
\frac{1}{2}\frac{\ddot a_0}{a_0}(t - t_0)^2 + ...\biggl] \quad .
\end{eqnarray}
Rappelons les d\'efinitions de la
 constante de Hubble et du
param\`etre de d\'ec\'el\'eration:
\begin{equation}
H_0 = \frac{\dot a_0}{a_0} \quad , \quad q_0 = - \frac{\ddot
a_0a_0}{\dot a_0^2} \quad .
\end{equation}
En utilisant la d\'efinition du d\'ecalage vers le rouge $z$, nous
obtenons
\begin{equation}
\frac{1}{1 + z} = 1 + H_0(t - t_0) - \frac{1}{2}q_0H_0^2(t -
t_0)^2 + ... \quad .
\end{equation}
En inversant, on obtient:
\begin{equation}
z = H_0(t_0 - t) + \biggr(1 + \frac{q_0}{2}\biggl)H_0^2(t_0 - t)^2
 +... \quad .
 \end{equation}
 Maintenant, nous faisons un d\'evelopement limit\'e de la relation
 \begin{equation}
 r = \int_{t_i}^{t_0}\frac{dt}{a(t)}  \quad ,
 \end{equation}
 obtenant
 \begin{equation}
 r = (t_0 - t) + \frac{1}{2}H_0(t_0 - t)^2 + ... \quad .
 \end{equation}
 conduisant \`a
 \begin{equation}
 \label{r1}
 rH_0 = z - \frac{1}{2}(1 + q_0)z^2 + ... \quad .
 \end{equation}
 Remarquons que le premier terme nous fournit la loi de Hubble,
 puisque le d\'ecalage vers le rouge $z$ est directement reli\'e
 \`a la vitesse de la source. L'effet Doppler relativiste s'\'ecrit comme
\begin{equation}
\nu = \nu_0\frac{\sqrt{1 - \frac{v^2}{c^2}}}{1 - \frac{v_r}{c}}
\quad ,
\end{equation}
o\`u $\nu$ est la fr\'equence observ\'ee, $\nu_0$ est la
fr\'equence dans le rep\`ere propre de la source, $v$ est la
vitesse de la source par rapport \`a l'observateur et $v_r$ est la
vitesse radiale de la source.
 En fait, lorsque la vitesse de la
 source est petite par rapport \`a la vitesse de la lumi\`ere, et
 qu'elle se d\'eplace radialement, nous avons
 \begin{equation}
 z = \frac{v}{c} \quad ,
 \end{equation}
 d'o\`u, au premier ordre,
 \begin{equation}
\label{lh}
 v = H_0r \quad ,
 \end{equation}
 qui est la loi de Hubble reliant la vitesse de r\'ecession avec
 la distance de la source.
 \par
 Un concept tr\`es important est celui de la {\it
 distance luminosit\'e}. Consid\'erons une source avec une
 luminosit\'e absolue $L£$, c'est-\`a-dire, $L$ est la quantit\'e
 d'\'energie \'emise par la source par unit\'e de temps et unit\'e d'aire.
 Avec la normalisation $a_0 = 1$, nous aurions que la luminosit\'e
 re\c{c}u devrait \^etre de l'ordre
 \begin{equation}
 l = \frac{L}{4\pi r^2} \quad ,
 \end{equation}
 $r$ etant la distance de la source aux observateurs. Mais, \`a cause de
l'expansion de l'Univers, l'\'energie de chaque photon \'emis par
 la source d\'ecro\^{\i}t par un facteur $a_1/a_0 = a_1$, o\`u $a_1$
 est la valeur du facteur d'\'echelle au moment o\`u le photon a
 \'et\'e \'emis; par contre, la quantit\'e de photons re\c{c}us
 d\'ecro\^{\i}t aussi du m\^eme facteur \`a cause de l'augmentation de
 la p\'eriode d'\'emission. Ainsi,
 nous avons pour la luminosit\'e mesur\'ee par l'observateur,
 \begin{equation}
 l = \frac{a^2(t)L}{4\pi r^2} \quad ,
 \end{equation}
 $a(t)$ \'etant le facteur d'\'echelle au moment de l'\'emission et
 $r$ \'etant la distance de la source \`a l'observateur.
 Ainsi, nous pouvons d\'efinir la distance luminosit\'e comme
 \begin{equation}
 d_L = \frac{r}{a} \quad .
 \end{equation}
\par
En astronomie on d\'efinit la magnitude apparente $m$ et la
magnitude absolue $M$ par la relation
\begin{equation}
d_L = 10^{1 + \frac{m - M}{5}} \quad .
\end{equation}
Mais, nous avons pour la distance luminosit\'e, en utilisant le
concept de d\'ecalage vers le rouge employ\'e auparavant,
\begin{equation}
d_L = r(1 + z) \quad .
\end{equation}
En utilisant l'expression (\ref{r1}), nous obtenons
\begin{equation}
d_L = \frac{1}{H_0}\biggr[z + \frac{1}{2}(1 - q_0) + ...\biggl]
\quad .
\end{equation}
Donc, en mesurant la distance luminosit\'e d'une source dont on
conna\^{\i}t la magnitude absolue, on peut essayer de d\'ecouvrir
$h_0$ et $q_0$. Ayant ces quantit\'es on peut obtenir, \`a l'aide
les \'equations d'Einstein, des informations sur le contenu
mat\'eriel de l'Univers.

 \section{Le mod\`ele cosmologique standard}

 La cosmologie standard, qui essaye de d\'ecrire l'Univers que l'on
 observe, est bas\'ee sur un nombre limit\'e des faits
 observationnels:
 \begin{enumerate}
 \item L'Univers est en expansion;
 \item L'Univers aujourd'hui a une temp\'erature isotrope de
 l'ordre de $2,7\,K$;
 \item La distribution de temp\'erature dans le ciel pr\'esente une
 petite anisotropie de l'ordre de
 \begin{equation}
 \frac{\Delta T}{T} \sim 10^{-5} \quad ;
 \end{equation}
 \item La mati\`ere baryonique est compos\'ee, en gros, de
 $75\%$ d'hydrog\`ene et $25\%$ d'h\'elium;
 \item Il existe des structures locales dans l'Univers sous
 forme de galaxies, amas de galaxies, amas d'amas de galaxies,
 r\'egions vides, la grande paroi, etc. La densit\'e dans le cas
 d'une agglom\'erations de mati\`ere peut atteindre $100$ fois la
 valeur de la densit\'e moyenne de l'Univers.
 \end{enumerate}
Il y a un autre r\'esultat, assez r\'ecent, venant des observations
des supernovas type Ia, qui indique que l'Univers se trouve dans
une phase d'expansion acc\'el\'er\'ee. Mais, nous aborderons cette
question plus tard.
\par
L'expansion de l'Univers ressort des solutions trouv\'ees
auparavant. Consid\'erant un fluide avec une pression $p =
\alpha\rho$, nous avons trouv\'e des solutions o\`u $\dot a$ est
toujours positif, sauf dans le cas o\`u la courbure est $k = 1$:
pour cet Univers, dit Univers ferm\'e, \`a parti d'un certain
moment, $\dot a$ peut \^etre negatif. Mais, en tout cas, le fait
que l'Univers soit en expansion aujourd'hui, et depuis longtemps
il faut le dire, n'est pas en contradiction avec un Univers
ferm\'e.
\par
L'\'evidence que l'Univers est en expansion vient de
l'observation des galaxies autour de la notre: \`a part certaines
galaxies qui appartiennent \`a notre groupe local de galaxies et
qui, par cons\'equent, ont des vitesses peculi\`eres (vitesses propres par
rapport au flot cosmologique) tr\`es
importantes, toutes les autres semblent s'\'ecarter de nous. Si
nous n'occupons pas une place privil\'egi\'ee dans l'Univers, cela
indique que des observateurs dans d'autres galaxies verraient le
m\^eme sc\'enario. Donc, la seule possibilit\'e c'est que
l'Univers doit se trouver en expansion. La vitesse d'expansion
peut \^etre mesur\'ee \`a partir du d\'ecalage vers le rouge des
raies spectrales observ\'ees dans le rayonnement re\c{c}u de ces
objets. La distance, du moins des galaxies proches, peut \^etre
mesur\'ee par des moyens ind\'ependants (par exemple, gr\^ace aux
\'etoiles variables du type C\'eph\'e\"{\i}des). Ainsi, nous pouvons obtenir
la valeur de la constante de Hubble aujourd'hui en utilisant
l'expression (\ref{lh}). Des mesures r\'ecentes indiquent
\begin{equation}
H_0 \sim 70\frac{km}{Mpc.s} \quad .
\end{equation}
Puisque $1\,Mpc \sim 3\times10^{24}\,cm$, cela nous conduit \`a
$H_0 \sim 4\times10^{17}\,s^{-1}$. Nous avons affirm\'e auparavant que
l'\^age de l'Univers est de l'ordre de l'inverse de la constante
de Hubble. Ainsi nous obtenons que l'Univers doit avoir un \^age
de l'ordre de
\begin{equation}
t_0 \sim 13\,Ga,
\end{equation}
o\`u $1\,Ga = 10^{9}$ ans. Il faut dire que l'\^age estim\'e des
objets les plus vieux qu'on connait dans l'Univers, les amas
globulaires d'etoiles, est compris entre $11$ et $16\,Ga$. Ces
estimations \'etant tr\`es impr\'ecises, il n'est pas clair pour
l'instant s'il existe une "crise d'\^age" ou pas, c'est-\`a-dire,
si l'\^age des enfants (les amas globulaires) est plus \'elev\'e
que l'\^age des parents (l'Univers lui-m\^eme).
\par
En tout cas, ayant l'\^age de l'Univers, nous pouvons estimer "son
rayon visible" en le multipliant par la vitesse de la lumi\`ere:
\begin{equation}
R_H = \frac{c}{H_0} \sim 4.000\,Mpc \sim 10^{28}\, cm \quad .
\end{equation}
Maintenant, poss\'edant la valeur de la constante de Hubble
aujourd'hui on peut estimer la densit\'e critique: celle pour que
l'Univers soit spatialement plat. Nous avons vu que
\begin{equation} H_0^2 = \frac{8\pi}{3}
G\sum_{i=1}^n\rho_{i0} = \frac{8\pi}{3} G\rho_c \quad.
\end{equation}
En ins\'erant les valeurs num\'eriques, on obtient pour la
densit\'e critique de l'Univers, $\rho_c \sim 10^{-28}\,g/cm^3$.
Les observations directes du contenu mat\'eriel de l'Univers
identifient deux composantes principales: la mati\`ere baryonique
et le rayonement. Les baryons ont une \'equation d'\'etat $p_b = 0$,
tandis que le rayonement, le gaz de photon responsable de la
temp\'erature de l'Univers, ob\'eit \`a l'\'equation d'\'etat $p_r =
\rho_r/3$. Nous avons d\'efini auparavant les param\`etres de
masse en fonction de la densit\'e critique, $\Omega_{i0}$. Les
observations indiquent
\begin{equation}
\Omega_{b0} \sim 0,04 \quad , \quad \Omega_{r0} \sim 10^{-5} \quad
.
\end{equation}
Donc, la mati\`ere baryonique domine aujourd'hui sur le
rayonnement. Mais, \`a cause des \'equations d'\'etat de chacune de ces
composantes, leur densit\'e \'evolue avec le facteur d'\'echelle, ou
avec le d\'ecalage vers le rouge $z$, comme
\begin{equation}
\Omega_b = \frac{\Omega_{b0}}{a^3} = \Omega_{b0}(1 + z)^3 \quad ,
\quad \Omega_r = \frac{\Omega_{r0}}{a^4} = \Omega_{r0}(1 + z)^4
\quad .
\end{equation}
Puisque la densit\'e du gaz photonique d\'ecro\^{\i}t plus vite que
celle des baryons dans un Univers en expansion, il doit y avoir dans le
pass\'e un moment o\`u les deux densit\'es doivent
\^etre \'egales. En fait, en posant $\Omega_b = \Omega_b$, on
trouve que cela a eu lieu vers $z = 4.000$. Si l'on consid\`ere un
Univers plat, domin\'e par la mati\`ere, nous avons $a \propto
t^{2/3}$, et cette valeur de $z$, compte tenu de l'\^age de
l'Univers estim\'e ci-dessus, \'equivaut au moment o\`u l'Univers
avait trois cents millions d'ann\'ees: jusqu'\`a ce moment, le
rayonnement dominait sur la mati\`ere baryonique. Remarquons que
l'estimation de la valeur de $z$ quand il y a eu \'egalit\'e
entre la mati\`ere et le rayonnement est, grosso modo, ind\'ependante
du mod\`ele, tandis que l'estimation de l'\^age de l'Univers \`a
ce moment d\'epend fortement du mod\`ele (Univers plat, ferm\'e,
ouvert, etc.).
\par
Le mod\`ele standard de la cosmologie contient donc deux phases
principales, en principe. Dans un premier temps, la radiation domine le
contenu de l'Univers, jusqu'\`a $z = 4.000$; apr\`es, la mati\`ere
baryonique domine, jusqu'\`a aujourd'hui. Quelles sont les
\'evidences pour ce sc\'enario? En ce qui concerne la phase de
rayonnement, d'abord il y a la temp\'erature de l'Univers
aujourd'hui, un fossile de cette phase "chaude" de l'\'evolution
cosmique. Mais, il y a une autre \'evidence: la nucleosynth\`ese
primordiale. En fait, des \'el\'ements chimiques l\'egers, tels l'hydrog\`ene, le deut\'erium, l'h\'elium
et le lythium ont d\^u \^etre
produits dans l'Univers primordial. Des consid\'erations sur
l'Univers primitif montrent que ces \'el\'ements ont du \^etre
produits dans les $3$ premi\`eres minutes de
l'Univers. Des calculs d\'etaill\'es ont permis d'estimer
l'abondance des ces \'elements, et ces calculs fournissent des
r\'esultats qui sont en accord avec l'observation \`a $1\%$ pr\`es.
Cet accord remarquable est obtenu si le param\`etre de densit\'e
pour les baryons est tel qu'aujourd'hui $\Omega_{b0} \sim 0,04$
et que le rapport entre baryons et photons est $\eta = 10^{-10}$.
\par
En plus, la phase mat\'erielle est n\'ecessaire pour que des
structures comme les galaxies se forment. En fait, la mati\`ere
baryonique a une pression nulle, et lorsqu'elle domine, le
ph\'enom\`ene d'effondrement gravitationnel peut avoir lieu, ce
qui est \`a l'origine des structures locales. N\'eanmoins, ce
sc\'enario pr\'esente des difficult\'es qui sugg\`erent que
d'autres ingr\'edients doivent \^etre ajout\'es au mod\`ele
standard.

\section{Le probl\`eme de formation des structures}

Les galaxies, amas de galaxies et d'autres structures plus
complexes observ\'ees dans l'Univers doivent \^etre form\'ees \`a
partir de la mati\`ere \`a pression nulle, en l'ocurrence la
mati\`ere baryonique. Mais, lorsqu'on fait une \'etude
detaill\'ee du processus de formation des ces structures, on peut
se rendre compte que le sc\'enario esquiss\'e auparavant ne peut
expliquer convenablement le processus de formation des structures
de l'Univers. On fournira par la suite des raisons \`a la fois
qualitative et quantitative \`a  ce probl\`eme.
\par
Tout d'abord, remarquons que pendant la phase de rayonnement, la
mati\`ere baryonique ne peut subir le processus d'effondrement
gravitationnel. Cela tout simplement parce que, \`a ce moment, les
baryons sont fortement coupl\'es aux photons, et le gaz de photons
exhibe une pression tr\`es \'elev\'ee: cette pression photonique
emp\^eche les baryons de s'agglom\'erer. Donc, les baryons ne peuvent
s'agglom\'erer qu'\`a partir du moment o\`u ils commencent \`a dominer le
contenu mat\'eriel de l'Univers, c'est-\`a-dire, \`a partir de $z
= 4.000$. En fait, ils s'agglom\`erent \`a partir d'un moment
ult\'erieur, puisque, m\^eme si les photons ne dominent plus le
contenu mat\'eriel de l'Univers, il y a tellement plus de photons
que de baryons que ceux derniers restent encore tr\`es coupl\'es
aux photons jusqu'\`a $z \sim 1.000$. A partir de ce moment, les
baryons peuvent subir le processus d'effondrement gravitationnel.
\par
Pour voir comment a lieu ce processus, nous retournons au
formalisme newtonien. Cela rendra l'analyse plus simple, sans
sacrifier l'essentiel des r\'esultats. Retournons aux \'equations
(\ref{cn1},\ref{cn2},\ref{cn3}). Nous connaissons d\'ej\`a les
solutons qui d\'ecrivent l'\'evolution de l'Univers de base.
Maintenant, introduisons dans ces \'equations les quantit\'es,
\begin{equation}
\rho = \rho_0 + \rho_1 \quad , \quad \vec v = \vec v_0 + \vec v_1
\quad , \quad \phi = \phi_0 + \phi_1 \quad ,
\end{equation}
o\`u les quantit\'es avec les indices $0$ indiquent la solution
de fond, repr\'esentant l'Univers homog\`ene et isotrope non
perturb\'e, et les quantit\'es avec les indices $1$
repr\'esentent des petites fluctuations autour de ces solutions de
base. En g\'en\'eral, ces fluctuations ont la forme,
\begin{equation}
f(t,\vec x) = f(t)\exp\biggr[i\frac{\vec k.\vec x}{a}\biggl] \quad
\end{equation}
Le facteur d'\'echelle a \'et\'e introduit de sorte que $\vec k =
(2\pi/\lambda)\hat k$, $\lambda$ est la longueur d'onde, et
$a\lambda$ fournit la longueur d'onde physique. En introduisant
les fluctuations de cette mani\`ere, et gardant les termes
du premier ordre, nous obtenons lorsque la pression est
nulle,
\begin{eqnarray}
\dot\rho_1 + 3\frac{\dot a}{a}\rho_1 + i\frac{\rho_0}{a}\vec
k.\vec v_1 = 0 \quad , \\
\dot{\vec v}_1 + \frac{\dot a}{a}\vec v_1 = - i\frac{\vec k}{a}\frac{p_1}{\rho_0} - i\frac{\vec
k}{a}\phi_1
\quad , \\
- \frac{k^2}{a^2}\phi_1 = 4\pi G\rho_1 \quad .
\end{eqnarray}
En se rappellant que
\begin{equation}
\dot\rho_0 + 3\frac{\dot a}{a}\rho_0 = 0 \quad ,
\end{equation}
et en d\'efinissant le contraste de densit\'e
\begin{equation}
\Delta = \frac{\rho_1}{\rho_0} \quad ,
\end{equation}
nous obtenons
\begin{equation}
\label{ep}
\ddot\Delta + 2\frac{\dot a}{a}\dot\Delta + \biggr[\frac{k^2v^2_s}{a^2} - \frac{3}{2}\biggr(\frac{\dot
a}{a}\biggl)^2\biggl]\Delta = 0 \quad,
\end{equation}
o\`u nous avons defini
\begin{equation}
v^2_s = \frac{\partial p}{\partial\rho} \quad ,
\end{equation}
la vitesse du son. Le dernier terme dans l'\'equation (\ref{ep}) montre la competition entre deux effets:
le terme connect\'e \`a la vitesse du son, et le terme connect\'e \`a la masse.
Lorsque le terme de la vitesse du son (donc, de la pression) domine, les perturbations oscillent,
tandis que quand le terme gravitationnel domine, les perturbations cro\^{\i}ent. Cela conduit au concept de
{\it longueur d'onde de Jeans}, la
longueur d'onde assoaci\'e \`a une fluctuation \`a partir de laquelle l'enffondrement gravitationnel a lieu. Puisque
\begin{equation}
\frac{3}{2}\biggr(\frac{\dot
a}{a}\biggl)^2 = 4\pi G\rho \quad ,
\end{equation}
dans le cas
pr\'esent, la longueur d'onde de Jeans prend la forme
\begin{equation}
\lambda_j = \frac{\sqrt{\pi}v_s}{a\sqrt{G\rho}} \quad .
\end{equation}
Dans ce qui suit, la vitesse du son sera \'egale \`a z\'ero,
puisque nous avons \`a faire \`a de baryons.
Pour obtenir cette \'equation nous avons \'egalement utilis\'ee
l'\'equation de Friedmann pour le cas plat.
\par
Maintenant, rappelons que pour le cas plat $a \propto t^{2/3}$.
Ainsi, nous avons l'\'equation
\begin{equation}
\ddot\Delta + \frac{4}{3}\frac{\dot\Delta}{t} -
\frac{2}{3}\frac{\Delta}{t^2} = 0 \quad .
\end{equation}
Nous avons deux solutions:
\begin{equation}
\Delta_+ \propto t^{2/3} \quad , \quad \Delta_- \propto t^{-1}
\quad .
\end{equation}
La deuxi\`eme solution repr\'esente un mode d\'ecro\^{\i}ssant, et
en principe il ne nous int\'eresse pas pour ce qui concerne la
formation des structures. La premi\`ere solution repr\'esente un
mode cro\^{\i}ssant, et il se comporte comme $\Delta \propto a$,
c'est-\`a-dire, il cro\^{\i}t comme le facteur d'\'echelle.
\par
Maintenant nous sommes en mesure de v\'erifier si le sc\'enario
standard peut rendre compte du processus de formation de galaxies.
Cependant, les galaxies sont des objets qui peuvent parfois avoir une
densit\'e de deux ordres de grandeurs plus grande que la densit\'e
critique. C'est-\`a-dire que les galaxies sont dans le r\'egime non
lin\'eaire. Mais, les mesures de l'anisotropie du rayonnement
cosmique de fond indiquent que les fluctuations de temp\'erature au
moment o\`u la radiation s'est d\'ecoupl\'ee de la mati\`ere \'etaient de
l'ordre de $10^{-5}$. Puisque les deux composantes, mati\`ere et
radiation, \'etaient coupl\'ees \`a ce moment, on peut s'attendre
\`a ce que cela soit l'amplitude du contraste de densit\'e.
Depuis, le contraste de densit\'e a \'evolu\'e comme le facteur
d'\'echelle. Donc nous avons que le contraste de densit\'e
aujourd'hui doit \^etre de l'ordre de
\begin{equation}
\Delta_f = \frac{a_f}{a_i}\Delta_i \quad .
\end{equation}
Posant, $a_f = a_0 = 1$, $a_i = 1/(1 + z_i) \sim 10^{-3}$ et
$\Delta_i \sim 10^{-5}$, nous obtenons pour le contraste de
densit\'e aujourd'hui,
\begin{equation}
\Delta_f = \Delta_0 \sim 10^{-2} \quad .
\end{equation}
Ainsi, avec le sc\'enario d\'ecrit ci-dessus, nous serions loin
encore du r\'egime non lin\'eaire, et les galaxies n'auraient pas
encore \'et\'e form\'ees.
\par
Il y a une mani\`ere de r\'esoudre ce probl\`eme. Elle
consiste \`a supposer qu'en plus de baryons il existe dans
l'Univers une autre type de mati\`ere froide, c'est-\`a-dire sans
pression, dont l'interaction avec le rayonnement et les baryons est
tr\`es faible. Ce genre de mati\`ere est appell\'e {\it WIMPS},
de l'anglais {\it weak interacting massive particles}, particules
massives d'interaction faible. Cette mati\`ere,
justement \`a cause de sa faible interaction, se d\'ecouple des
photons bien avant les baryons. Ainsi, elle peut
commencer \`a subir le processus d'effondrement gravitationnel
bien avant les baryons, et atteindre le r\'egime non lin\'eaire
avant aujourd'hui. En plus, l'agglom\'eration de ces {\it WIMPS}
cr\'ee des puits de potentiels o\`u les baryons tombent apr\`es,
ce qui explique pourquoi on trouve des baryons aujourd'hui dans le
r\'egime non lin\'eaire.
\par
Il existe uniquement un inconv\'enient \`a ce sc\'enario: ces
particules massives n'ont jamais \'et\'e d\'etect\'ees. Elles ne
peuvent \^etre form\'ees de quarks et de leptons comme les baryons,
puisque sinon elles devraient \'emettre de la radiation; puisque cette
mati\`ere n'\'emet aucun type de radiation on l'appelle
mati\`ere noire. Il existent des candidats pour cette mati\`ere
froide noire, comme les axions, qui sont pr\'evus par des
th\'eories de grande unification. Mais cela reste sp\'eculatif. De plus, l'\'etude de la dynamique des amas de galaxies
indiquent un param\`etre de densit\'e pour la mati\`ere sans
pression de l'ordre de
\begin{equation}
\Omega_{m0} \sim 0,2 \quad ,
\end{equation}
bien au-dessus de la quantit\'e de baryons pr\'evue par la
nucl\'eosynth\`ese primordiale, et en accord avec le sc\'enario de
mati\`ere froide noire pour la formation de galaxies. Cela  soutient
cette id\'ee d'une autre composante non-baryonique dans l'Univers.

\section{Le probl\`eme de conditions initiales et le sc\'enario
inflationaire}

Le mod\`ele cosmologique standard \'ebauch\'e pr\'ec\'edement contient
plusieurs difficult\'es qui doivent \^etre abord\'ees par un
mod\`ele cosmologique complet. Les probl\`emes les plus remarquables,
\`a par l'existence d'une singularit\'e initiale, ont \`a voir
avec la n\'ecessit\'e d'imposer des conditions initales tr\`es
particuli\`eres. Ces conditions initiales concernent:
\begin{enumerate}
\item Le probl\`eme de la platitude de l'Univers;
\item Le
probl\`eme de l'horizon;
 \item Le probl\`eme du spectre initial
des perturbations cosmologiques.
\end{enumerate}
\par
Commen\c{c}ons par le probl\`eme de la platitude. A un temps $t$
quelconque, nous pouvons \'ecrire les \'equations d'Einstein comme
\begin{equation}
H^2 + \frac{k}{a^2} = \frac{8\pi G}{3}\rho_T \quad .
\end{equation}
Cette \'equation peut \^etre r\'e-\'ecrite comme
\begin{equation}
\Omega_T - 1 = \frac{k}{a^2H^2} = \frac{k}{\dot a^2} \quad .
\end{equation}
Lorsque l'Univers est compos\'e des fluides "normaux" dont la
pression est positive, il est en train de d\'ec\'el\'erer. Dans ce
cas $\dot a$ est une fonction d\'ecroissante du temps. Ainsi, pour
une valeur de $k$ donn\'ee, la difference $|\Omega_T - 1|$ a d\^u
\^etre beaucoup plus petite dans le pass\'e qu'aujourd'hui. Or,
les observations concernant le spectre de l'anisotropie du
rayonnement de fond cosmique indiquent que, aujourd'hui
$|\Omega_{T0} - 1| \sim 0,02$. Consid\'erons, pour simplifier, le
cas d'un Univers plat domin\'e par le rayonnement pendant toute son
histoire. Par rapport \`a un calcul que tient en compte toutes les
phases de l'histoire de l'Univers, cela revient \`a introduire un
facteur de l'ordre de l'unit\'e. Ainsi, $a = a_0t^{1/2}$. Donc,
\begin{equation}
\Omega_T - 1 = \frac{4kt}{a_0^2} \quad .
\end{equation}
Comparons cette expression aujourd'hui et dans un temps recul\'e
dans le pass\'e:
\begin{equation}
\frac{\Omega_T - 1}{\Omega_{T0} - 1} = \frac{t}{t_0} \quad .
\end{equation}
Ainsi, en introduisant $|\Omega_{T0} - 1| \sim 0,02$, $t_0 \sim
10^{17}\,s$ (l'\^age de l'Univers aujourd'hui) et, par example, $t
= 10^{11}\,s$ (a peu pr\`es le moment de l'egalit\'e entre le
rayonement et la mati\`ere), nous obtenos
\begin{equation}
|\Omega_T - 1| \sim 10^{-6} \quad .
\end{equation}
Il est facile de v\'erifier que plus nous reculons dans le
pass\'e, plus $\Omega_T$ est proche de l'unit\'e,
c'est-\`a-dire de la densit\'e critique. Ceci est d\^u au fait
que $\Omega_T = 1$ (la densit\'e critique) est un point instable:
si aujourd'hui il est proche de l'unit\'e, comme indiquent les
observations, il devrait \^etre extremement proche de l'unit\'e
dans le pass\'e tr\`es lointain. Ceci est une condition initiale
tr\`es particuli\`ere. La question est: pourquoi cette condition a
eu lieu dans l'Univers primordial?
\par
Le probl\`eme de l'horizon peut \^etre \'enonc\'e comme suit.
Consid\'erons, de nouveau pour simplifier les calculs, que
l'Univers est plat et domin\'e par le fluide de radiation, comme avant. Nous savons que
l'Univers est rempli par une radiation isotrope; cela veut dire
que la temperature est la m\^eme dans toutes les directions. Or,
depuis le d\'ecouplage entre la radiation et la mati\`ere, les
photons voyagent librement dans l'espace. Consid\'erons de nouveau
pour simplifier, que le d\'ecouplage a eu lieu \`a $t_d \sim
10^{11}\,s$. Puisque nous avons affaire \`a la propagation de
photons, la distance coordonn\'ee d'une source qui a \'emis le
photon qui nous arrive aujourd'hui venant de la "surface de derni\`ere difusion" (la surface d\'efinie par le moment du
d\'ecouplage entre la radiation et la mati\`ere), nous avons
\begin{equation}
ds^2 = 0 \quad \rightarrow \quad r = \int_{t_d}^{t_0}\frac{dt}{a}
= \int_{t_d}^{t_0}\frac{dt}{a_0t^{1/2}} = 2\biggr(t_0^{1/2} -
t_d^{1/2}\biggl) \sim 2t_0^{1/2} \quad ,
\end{equation}
puisque $t_0 \sim 10^{17}\,s$. Si nous consid\'erons deux photons
qui ont \'et\'e emis par des sources diam\'etralement oppos\'ees,
la distance coordon\'ee entre ces deux sources est
\begin{equation}
r_s = 2r = \sim 4t_0^{1/2} \quad .
\end{equation}
Observons que ces sources pr\'esentent la m\^eme temp\'erature.
\par
On calcule maintenant la distance horizon coordonn\'ee au moment
du d\'ecouplage entre la radiation et la mati\`ere. La distance
horizon fournie la dimension de la r\'egion causale \`a ce
moment. Elle est obtenue en cherchant la distance (toujours
coordonn\'ee) parcourue par la lumi\`ere de $t = 0$ jusqu'au
moment qu'il nous int\'eresse, c'est-\`a-dire, le moment du
d\'ecouplage:
\begin{equation}
r_H = \int_0^{t_d}\frac{dt}{a} = \int_0^{t_d}\frac{dt}{a_0t^{1/2}}
= \frac{2t_d^{1/2}}{a_0} \quad .
\end{equation}
Divisons maintenant la distance coordonn\'ee qui s\'epare les
sources calcul\'e auparavant, et la distance coordonn\'ee de
l'horizon au moment du d\'ecouplage. Nous obtenons,
\begin{equation}
\frac{r_s}{r_H} \sim 2\frac{t_0^{1/2}}{t_d^{1/2}} \sim 200 \quad .
\end{equation}
Cela veut dire que les sources \'etaient s\'epar\'ees par une
distance \`a peu pr\`es deux cent fois plus grande que la distance
de connexion causale. Donc, les sources n'\'etaient pas en contact
causal. Pourtant, elles sont \`a la m\^eme temp\'erature. Ceci ne
peut \^etre obtenu qu'imposant l'equilibre thermique entre des
r\'egions sans connexion causale comme condition initiale.
\par
Ces probl\`emes peuvent \^etre r\'esolus si l'on suppose que, bien
au d\'ebut de l'histoire de l'Univers, il y a eu une p\'eriode
d'expansion exponentielle, ou quasi-exponentielle, dite p\'eriode
inflationnaire. Ceci peut avoir lieu dans un temps tr\`es petit.
Le lecteur peut consulter les r\'ef\'erences \cite{guth,liddle} pour plus de d\'etails.
Supposons donc que l'Univers subi une p\'eriode d'expansion exponentielle
entre en temps $t_i$ et $t_f$ peu apr\`es $t = 0$. Revenons tout
d'abord au probl\`eme de la platitude. Supposons en plus que $t_i
= 10^{-33}\,s$ pour des raisons dont on parlera plus tard. \`A
cette \'epoque, suivant le raisonement expos\'e ci-dessus, pour
expliquer les donn\'ees observationnelles, nous devrions avoir,
\begin{equation}
|\Omega_M - 1| \sim 10^{-50}
\end{equation}
Ce r\'esultat a \'et\'e obtenu en supposant que le facteur
d'\'echelle \'evoluait comme dans la phase de rayonnement. Si nous
supposons maintenant qu'il a exist\'e une phase exponentielle
comme d\'ecrite ci-dessus, pendant laquelle le facteur d'\'echelle
se comporte comme $a \propto \exp(\Gamma t)$, o\`u $\Gamma$ est
une constante, nous aurions \`a la fin de la phase exponentielle
\begin{equation}
\frac{|\Omega_M - 1|_f}{|\Omega_M - 1|_i} =
\exp[-2\Gamma(t_f-f_i)] \quad .
\end{equation}
Or, si $\Gamma\Delta t \sim 25$, l'Univers pourrait d\'ebuter
avec des valeurs pour le param\`etre de densit\'e de l'ordre de
l'unit\'e, et nous aurions m\^eme ainsi les valeurs requises par
l'observation aujourd'hui.
\par
Cette phase inflationnaire pourrait aussi r\'esoudre la question
de l'horizon. En fait, supposons qu'entre $t_i$ et $t_f$ le
facteur d'\'echelle exhibe une expansion exponentielle, et apr\`es
une expansion comme pour l'Univers domin\'e par le rayonnement.
Posons:
\begin{equation}
a = a_1e^{\Gamma t} \quad , \quad (t_i < t < t_f) \quad ; \quad a
= a_0t^{1/2} \quad , \quad (t_f < t) \quad .
\end{equation}
Les conditions de raccordement de ces solutions \`a $t = t_f$,
nous disent que
\begin{equation}
a_1 = a_0\frac{t_f^{1/2}}{\exp(\Gamma t_f)} \quad .
\end{equation}
Calculons de nouveau la distance horizon coordonn\'ee \`a la fin
de la phase inflationnaire:
\begin{equation}
r_H = - \frac{1}{\Gamma a_1}\biggr(e^{-\Gamma t_f} - e^{-\Gamma
t_i}\biggl) = - \frac{1}{\Gamma a_0t_f^{1/2}}\biggr(1 -
e^{\Gamma(t_f - t_i)}\biggl) \sim \frac{1}{\Gamma
a_0t_f^{1/2}}e^{\Gamma(t_f - t_i)} \quad ,
\end{equation}
si $\Gamma(t_f - t_i)$ est beaucoup plus grand que l'unit\'e. Si
l'on compare la distance coordonn\'ee de l'horizon \`a la fin de la phase
inflationnaire avec la distance coordonn\'ee entre les sources que nous observons aujourd'hui, nous avons
\begin{equation}
\frac{r_s}{r_H} = 2\Gamma(t_0t_f)^{1/2}e^{-\Gamma(t_f-t_i)} \sim
2\Gamma\times10^{-8}\times e^{-\Gamma\Delta t} \quad . \quad .
\end{equation}
Ceci peut \^etre de nouveau de l'ordre de l'unit\'e si
$\Gamma\Delta t \sim 65$. En fait, en supposant que la dur\'ee
de la phase inflationnaire est telle que $\Delta t \sim
10^{-33}\,s$, nous avons $\Gamma \sim 6\times10^{34}$ et, par
cons\'equent, $r_s/r_H \sim 1$ d\'ej\`a \`a la fin de la phase
inflationnaire: les sources qui ont \'emis les photons au
d\'ecouplage etaient d\'ej\`a causuellement connect\'ees d\`es la
fin de la phase inflationnaire.
\par
Maintenant, la question est comment obtenir cette phase. Si nous
regardons l'\'equation de Friedmann pour le cas plat, il est
facile de v\'erifier que pour ce faire il faut que, pendant la
p\'eriode inflationnaire, un fluide avec densit\'e constante doit
dominer le contenu mat\'eriel de l'Univers. Ceci peut \^etre
obtenu si l'on ajoute aux \'equations d'Einstein une constante
cosmologique $\Lambda$. Dans le cas o\`u il n'y a pas de mati\`ere, les
\'equations d'Einstein s'\'ecrirait,
\begin{equation}
R_{\mu\nu} - \frac{1}{2}g_{\mu\nu}R - g_{\mu\nu}\Lambda = 0 \quad ,
\end{equation}
et les \'equations du mouvement se liraient,
\begin{eqnarray}
3\biggr(\frac{\dot a}{a}\biggl) = \Lambda \quad , \\
2\frac{\ddot a}{a} + \biggr(\frac{\dot a}{a}\biggl)^2 = \Lambda
\quad ,
\end{eqnarray}
avec la solution
\begin{equation}
a \propto e^{\sqrt{\frac{\Lambda}{3}}t} \quad .
\end{equation}
Ces \'equations \'equivalent \`a celles d'un fluide parfait
d\'ecrites auparavant, avec une \'equation d'\'etat
\begin{equation}
p = - \rho \quad .
\end{equation}
Cette \'equation d'\'etat peut repr\'esenter le vide quantique. On
pourrait interpr\'eter la phase inflationnaire comme celle o\`u
l'\'energie du vide quantique domine sur les autres formes
d'\'energie.
\par
Mais, il y a une question relative \`a l'existence de cette
phase inflationnaire: puisque tous les autres formes d'\'energie que
nous avons expos\'e auparavant decro\^{\i}ssent avec le temps, et
puisque la phase inflationnaire doit avoir eu lieu uniquement dans
une p\'eriode de temps tr\`es courte bien avant la
nucleosynth\`ese (\'epoque o\`u l'Univers doit \^etre domin\'e
par le rayonnement), une fois que l'\'energie (constante) du vide
commence \`a dominer le contenu mat\'eriel de l'Univers, pourquoi
elle cesse de le faire \`a partir d'un certain moment? Ce
probl\`eme est tellement important qu'il a
re\c{c}u un nom: {\it le probl\`eme de la sortie gracieuse}.
\par
Une mani\`ere de traiter l'inflation et le probl\`eme de la sortie
gracieuse est consid\'erer que l'inflation n'est pas caus\'e par
la constante cosmologique proprement dite, mais par un champ
scalaire dynamique, avec auto-interaction, que se comporte comme une
constante cosmologique \`a un certain moment. En fait,
consid\'erons un champ scalaire avec auto-interaction donn\'e par
le lagrangien,
\begin{equation}
L = \sqrt{-g}\biggr[\phi_{;\rho}\phi^{;\rho} - 2V(\phi)\biggl]
\quad,
\end{equation}
o\`u $V(\phi)$ est le potentiel repr\'esentant l'auto-interaction
du champ. Le tenseur d'impulsion-\'energie associ\'e \`a ce
lagrangien est
\begin{equation}
T_{\mu\nu} = \phi_{;\mu}\phi_{;\nu} -
\frac{1}{2}g_{\mu\nu}\phi_{;\rho}\phi^{;\rho} + g_{\mu\nu}V(\phi) \quad .
\end{equation}
Si l'on consid\`ere maintenant la m\'etrique de FRW et les
expressions correspondante pour un fluide parfait, la densit\'e
d'\'energie et la pression associ\'ees \`a ce champ scalaire sont
\begin{eqnarray}
\rho = \frac{\dot\phi^2}{2} + V(\phi) \quad , \\
p = \frac{\dot\phi^2}{2} - V(\phi) \quad .
\end{eqnarray}
Si le potentiel est tel qu'il domine, pendant un certain moment de
l'\'evolution de l'Univers, sur le terme cin\'etique, nous avons
\begin{equation}
\rho \approx V(\phi) \quad , \quad p \approx - V(\phi) \quad
\rightarrow \quad p \approx - \rho \quad .
\end{equation}
\par
Tout le travail maintenant consiste \`a trouver un potentiel
$V(\phi)$ qui puisse satisfaire cette condition dans une courte
p\'eriode de temps bien au d\'ebut de l'histoire de l'Univers,
conduisant apr\`es au mod\`ele standard d\'ecrit auparavant. Il
existe plusieurs propositions dans ce sens dans la
litt\'erature, mais jusqu'ici aucun sc\'enario enti\`erement
satisfaisant n'a pu \^etre mis en place. Mais, en g\'en\'eral des champs scalaires
avec auto-interaction appara\^{\i}ssent dans le processus de brisure de symm\'etrie
dans des th\'eories comme la th\'eorie de grande unification. Cette brisure de sym\'etrie
a lieu \`a des \'echelles d'\'energie tr\`es hautes ($T \sim 10^{15}\,Gev$, par exemple), de mani\`ere
que, en terme du temps cosmiques, elles ont lieu \`a des \'echelles de temps tr\`es petites, comme celles
consid\'er\'ee ci-dessus. D'autre part, le probl\`eme de la sortie de la phase inflationnaire d\'ecrit ci-essus
peut \^etre resolu en introduisant un couplage entre le champ d'inflaton et le terme d\'ecrivant le fluide
de rayonnement.
\par
En tout cas, l'existence de cette possible phase inflationnaire
aurait des cons\'equences tr\`es positives pour la compr\'ehension
de la formation des structures dans l'Univers. En fait, nous avons
touch\'e \`a ce probl\`eme avant, mais nous n'avons pas abord\'e
une question essentielle: quelles sont les conditions initiales
des fluctuations qui donnent naissance aux structures observ\'ees
dans l'Univers? L'existence d'une phase inflationnaire peut
r\'epondre \`a cette question. En effet, le champ scalaire
r\'eponsable pour l'inflation, qu'on appelle {\it inflaton}, doit
\^etre d'origine quantique. Il doit subir donc des fluctuations
quantiques. En consid\'erant les fluctuations de l'inflaton comme
\'etant des fluctuations du vide quantique, on peut fixer
l'amplitude de ces perturbations, d\^u aux conditions de
normalisation des modes quantiques, et le spectre, d\^u au
caract\'ere gaussien des ces fluctuations du vide. Il est
remarquable que cet ensemble de conditions initiales dict\'ees par
le mod\`ele inflationnaire soit en accord apparent avec le spectre
des anisotropies du rayonnement cosmique de fond.

\section{L'acc\'el\'eration de l'Univers et autres questions en cosmologie}

Parmi les param\`etres cosmologiques, il y a un consensus aujourd'hui que la valeur
de la constante d'Hubble doit \^etre autour de $H_0 \sim 70\,km/Mpc.s$. D'autre part,
la grande surprise dans ces derni\`eres ann\'ees est venue de la d\'etermination du
param\`etre de d\'ec\'el\'eration $q_0$. Comme on a vu auparavant, ce param\`etre appara\^{\i}t dans
la d\'eviation du r\'egime lin\'eaire de la loi de Hubble. Ainsi, pour mesurer $q_0$, if faut observer
des objets \`a des valeurs assez \'el\'ev\'ees de $z$. Le grand probl\`eme que se pose est d'avoir des
"chandelles standards" fiables de mani\`ere qu'on puisse conna\^{\i}tre leur magnitude absolue et, au m\^eme temps, qu'on
puisse les detecter \`a des grandes distances. L'utilisation des galaxies posent, \`a cet \'egard, des probl\`emes
puisque nous ne connaissons pas leur processus \'evolutif.
\par
La possibilit\'e d'avoir des "chandelles standards" fiables, capables d'\^etre d\'etect\'ees \`a des grandes
distances, a re\c{c}ue une forte impulsion avec les observations des supernov\ae\, de type Ia. Ces supernov\ae
sont cons\'equences de l'explosion d'une \'etoile naine blanche, lorsque elle d\'epasse la limite de Chandrasekhar
d\^ue \`a l'absortion de la masse d'une \'etoile massive dans un syst\`eme binaire. Puisque la limite de stabilit\'e
des naines blanches est atteinte \`a une masse tr\`es pr\'ecise (la limite de Chandrasekhar), les conditions de l'explosion
de ces \'etoiles sont tr\`es standard, et la magnitude absolue qu'elles peuvent atteindre est bien \'etablie. Le probl\`eme
concernant ces supernovae\, type Ia r\'eside dans leur raret\'e: on observe tr\`es peu de supernov\ae\, de type Ia. Mais en regardant
tr\`es loin dans l'espace, on peut d\'etecter une quantit\'e raisonable de ces objets. En plus, comme leur luminosit\'e est tr\`es
\'elev\'ee, on peut d\'etecter des supernov\ae\, de type Ia \`a des valeurs de $z$ suffisament \'elev\'ees pour pouvoir avoir des
informations sur la valeur de $q_0$. 
\par
Deux projets de d\'etection des supernov\ae\, de type Ia sont en cours depuis la moiti\'e des ann\'ees
$90$ \cite{riess,mutter}. La grande surprise est venu du fait que l'accord avec les observations exige une valeur n\'egative pour
$q_0$. Cela implique, d'apr\`es les \'equations d'Einstein, que l'Univers aujourd'hui doit \^etre domin\'e par
un fluide de pression n\'egative, ceci de telle mani\`ere que l'\'equation d'\'etat effective est de l'ordre
$\alpha = p_T/\rho_t \sim - 0,7$. Il faut croiser ces r\'esultats avec ceux venant de l'\'etude du spectre de l'anisotropie
du rayonnement cosmique de fond. En fait, ce spectre, exprim\'e en termes d'un d\'evelopement multipolaire, pr\'esente
un plateau pour des valeurs petites de l'ordre multipolaire, suivi d'une s\'erie de pics, dit pics Doppler. La position
du premier pic Doppler a un rapport directe avec la densit\'e total de l'Univers. La position de ce premier pic
Doppler est coh\'erente avec une densit\'e totale,
\begin{equation}
\Omega_{T0} = 1,02 \pm 0,02 \quad .
\end{equation}
Ainsi, l'Univers aujourd'hui est quasiment plat.
\par
Donc, l'analyse de toutes ces donn\'ees implique que nous avons pour le contenu mat\'eriel de l'Univers aujourd'hui
la distribution suivante:
\begin{equation}
\Omega_{b0} \sim 0,04 \quad , \quad \Omega_{m0} \sim 0,3 \quad , \quad \Omega_{c0} \sim 0,7 \quad .
\end{equation}
$\Omega_{b0}$ indique la fraction de baryons, tandis que $\Omega_{m0}$ est la fraction de mati\`ere noire et $\Omega_{c0}$ indique
la fraction d'\'energie noire. Cette derni\`ere serait le fluide avec pression negative responsable de l'acc\'el\'eration de l'Univers
et qui ne s'agglom\'ererait pas, c'est-\`a-dire qu'elle reste une composante homog\`ene ne subissant pas le processus d'effondrement
gravitationnel.   
\par
Un aspect curieux des resultats esquiss\'es ci-dessus est que nous vivons aujourd'hui dans une phase inflationaire. Donc, le
mod\`ele d'inflation originairement con\c{c}u pour r\'esoudre les probl\`emes de conditions initiales de l'Univers primordial
doit \^etre \'egalement appliqu\'e \`a l'Univers actuel. Pour la phase inflationaire dans l'Univers primordial nous avons
d\'ej\`a vu quels sont les candidats pour le champ qui engendre l'expansion ac\'el\'er\'ee de l'Univers. Pour la phase inflationaire
que se manifeste aujourd'hui, il existe plusieurs candidats, tous ayant faire face \`a diff\'erentes difficult\'es.
Le premier candidat est la constante cosmologique vue comme \'energie du vide quantique. Admettre que la constante cosmologique
a \'et\'e responsable pour la phase inflationnaire primordiale pose des grandes difficult\'es puisque cette phase primordiale
doit finir bien avant la nucl\'eosynth\`ese primordiale; mais, une constante cosmologique une fois dominant le contenu mat\'eriel
de l'Univers, restera la composante dominante par la suite, ce qui rendre la constante cosmologique une candidate improbable pour
la premi\`ere phase d'inflation. N\'eanmoins, cet emp\^echement est moins important pour la phase actuelle. De toute mani\`ere, la
valeur de la constante cosmologiques n\'ecessaire pour expliquer l'ac\'el\'eration pr\'esente de l'Univers devrait \^etre telle que
$\Omega_{\Lambda0} \sim 10^{-47}\,GeV^4$. Cette valeur est $120$ ordre de grandeur plus petite que la valeur pr\'evue pour l'\'energie
du vide dans l'Univers par la th\'eorie quantique du champ. Ce fait rendre tr\`es improbable la possibilit\'e que l'\'energie du vide soit
la constante cosmologique\cite{caroll}.
\par
En vue de cette situation, il a \'et\'e propos\'ee aussi pour l'Univers actuel un champ dynamique qui serait responsable 
de l'acc\'el\'eration de l'Univers aujourd'hui. \c{C}a serait un champ scalaire avec auto-interaction,
tout comme l'inflaton d\'ecrit auparavant, mais
dont le potentiel aurait une autre forme, de mani\`ere \`a que ce champ scalaire soit au d\'ebut une composante sous-dominante, devenant
plus tard la composante dominante de l'Univers. Il a \'et\'e d\'enom\'e {\it quintessence}\cite{stein}. Des potentiels avec les caract\'eristiques
requises peuvent \^etre obtenus \`a partir, par exemple, des th\'eories de supergravit\'e. N\'eanmoins, un ajustement fin 
des param\`etres se fait n\'ecessaire. Il y a d'autres propositions pour la description de l'\'energie noire, tels le gaz de Chaplygin\cite{pasquier},
le champ fant\^ome\cite{alguem}, etc. Mais, pour l'instant la question de qui est effectivement responsable de l'acc\'el\'eration de l'Univers
aujourd'hui reste ouverte.
\par
Il existe plusieurs autres sujets importants aujourd'hui en cosmologie, soit du point de vue th\'eorique, soit du point de vue
observationnele. Nous aborderons, pour clore cette s\'erie d'expos\'es sur la cosmologie uniquement la question de la singularit\'e
initiale. Cela constitue \'evidemment un probl\`eme puisqu'une singularit\'e repr\'esente, du point de vue math\'ematique, une
incompletude g\'eodesique, et du point de vue physique une situation o\`u toute description par des lois pr\'ecises est impossible.
Ainsi, il existe plusieurs propositions pour r\'esoudre le probl\`eme de la singularit\'e initiale. Une premi\`ere proposition consiste
\`a dire que, lorsque l'Univers s'approche de cette singularit\'e, des effets quantiques deviennent de plus en plus importants, de
telle mani\`ere qu'ils finiront pour \'eviter que l'Univers atteigne cette singularit\'e. Cette proposition, par ailleurs tr\`es
int\'eressante, a l'inconv\'enient majeur de qu'il n'existe pas encore de th\'eorie quantique de la gravitation consistante\cite{quantum}. Elle
reste, cependant, une sp\'eculation.
\par
Une voie int\'eressante pour r\'esoudre le probl\`eme de la singularit\'e tout comme les autres probl\`emes concernant la cosmologie
primordiale vient des th\'eories de cordes\cite{polchi,gasperini}. Les th\'eories de cordes sont bas\'ees sur l'id\'ee que les
\'el\'ements fondamentaux de
la nature sont des cordes qui se d\'eplacent dans l'espace-temps. Pour des raisons de coh\'erence, la dimension de l'espace-temps
doit \^etre \'egale \`a 10; les cordes elles m\^emes doivent avoir la dimension de la longueur de Planck, c'est-\`a-dire,
$10^{-33}\,cm$. L'action qui d\'ecrit la partie bosonique de cette th\'eorie est celle de Nambu-Goto;
\begin{equation}
A = \int d^{10}x \sqrt{G} \quad ,
\end{equation}
o\`u $G$ est le determinant de la m\'etrique \`a dix dimensions.
Un d\'eveloppement perturbatif conduit \`a l'action effective des cordes
\begin{equation}
A = \int d^{10}x\sqrt{g}e^{-\phi}\biggr[R + \phi_{;\rho}\phi^{;\rho} - \Psi_{;\rho}\Psi^{;\rho}\biggl] \quad ,
\end{equation}
o\`u $\phi$ est dit le champ de dilaton et $\Psi$ le champ d'axion.
Il existe beaucoup d'espoirs que la th\'eorie de cordes puisse conduire \`a un sc\'enario primordial o\`u les
probl\`emes du mod\`ele standard disparaissent. Pour l'instant, ceci reste un espoir, et peut \^etre faut il aller
au-del\`a de l'action effective \'ecrite ci-dessus pour obtenir des r\'esultats satisfaisants.
\newline
\vspace{0.5cm}
\newline
{\it Remerciements:} Je voudrais exprimer ma gratitude aux organisateurs de l'{\it Ecole Internationale sur les
structures g\'eom\'etriaues et applications} pour leur accueil chaleureux \`a Dakar. Je tiens \`a remercier \`a Charles Fran\c{c}ois, J\'er\^ome Martin et Patrick Peter pour leurs
remarques et critiques.


\begin{thebibliography}{90}
\bibitem{weinberg} S. Weinberg, {\bf Gravitation and cosmology}, Wiley, New York(1972);
\bibitem{bernstein} J. Bernstein, {\bf An introduction to cosmology}, Prentice Hall, New Jersey(1998);
\bibitem{rich} J. Rich, {\bf Fundamentals of cosmology}, Springer-Verlag, Berlin(2001);
\bibitem{guth} S.K. Blau et A.H. Guth, in {\bf 300 years of gravitation}. Editeurs: S.W. Hawking et W. Israel. Cambridge University
Press, Cambridge(1987);
\bibitem{liddle} A.R. Liddle et D.H. Lyth, {\bf Cosmological inflation and large-scale structure}, Cambridge University Press,
Cambridge(2000);
\bibitem{riess} A. Riess et al, Astron. J. {\bf 116}, 1009(1998);
\bibitem{mutter} S. Perlmutter et al, Astrophys. J. {\bf 517}, 565(1999);
\bibitem{caroll} S.M. Carroll, Living Rev. Rel. {\bf 4}, 1(2001);
\bibitem{stein} R.R. Caldwell, R. Dave et P.J. Steinhardt, Phys. Rev. Lett. {\bf 80}, 1582(1998);
\bibitem{pasquier} A. Kamenshchik, U. Moschella et V. Pasquier, Phys. Lett {\bf B511}, 265(2001);
\bibitem{alguem} V.B. Johri, Phys. Rev. {\bf D70}, 041303(2004);
\bibitem{quantum} A. Ashtekar, {\it Gravity and the quantum},gr-qc/0410054;
\bibitem{polchi} J. Polchinski, {\bf String theory}, volumes I et II, Cambridge University Press, Cambridge(1998);
\bibitem{gasperini} M. Gasperini et G. Veneziano, Phys. Rep. {\bf 373}, 1(2003).

\end{thebibliography}
\end{document}